\documentclass[12pt]{article}

\input{epsf}
\usepackage[pdftex]{graphicx}
\usepackage{amsmath}
\usepackage{cite}
\usepackage{color}
\usepackage{amsfonts}
\usepackage{amssymb}
\def\1ad{\mbox{\normalsize $^1$}}
\def\2ad{\mbox{\normalsize $^2$}}
\def\3ad{\mbox{\normalsize $^3$}}
\def\4ad{\mbox{\normalsize $^4$}}

\def\5ad{\mbox{\normalsize $^5$}}
\def\6ad{\mbox{\normalsize $^6$}}
\def\7ad{\mbox{\normalsize $^7$}}
\def\8ad{\mbox{\normalsize $^8$}}

\def\gev{\mbox{ GeV}}

\def\dj{\hbox{d\kern-0.347em \vrule width 0.3em height 1.252ex depth
-1.21ex \kern 0.051em}}

\def\Re{{\rm Re\,}}

\newcommand{\ra}{\rightarrow}

\newcommand{\mev}{\mbox{MeV}}

\newcommand{\be}{\begin{equation}}
\newcommand{\ee}{\end{equation}}
\newcommand{\ben}{\begin{equation*}}
\newcommand{\een}{\end{equation*}}
\newcommand{\ba}{\begin{eqnarray}}
\newcommand{\ea}{\end{eqnarray}}
\newcommand{\ban}{\begin{eqnarray*}}
\newcommand{\ean}{\end{eqnarray*}}
\newcommand{\brr}{\begin{array}}
\newcommand{\err}{\end{array}}
\newcommand{\bc}{\begin{center}}
\newcommand{\ec}{\end{center}}

\newcommand{\bea}{\begin{eqnarray}}
\newcommand{\eea}{\end{eqnarray}}
\newcommand{\bean}{\begin{eqnarray*}}
\newcommand{\eean}{\end{eqnarray*}}

\newcommand{\eg}{\mbox{\it e.g.~}}
\newcommand{\ie}{\mbox{\it i.e.~}}
\newcommand{\cf}{\mbox{\it c.f.~}}
\newcommand{\vev}[1]{\mbox{$\langle #1 \rangle $}}

\newcommand{\bk}{{\mathbf k}}

\newcommand{\bp}{{\mathbf p}}

\newcommand{\bx}{{\mathbf x}}

\newcommand{\bv}{{\mathbf v}}
\newcommand{\bq}{{\mathbf q}}

\newcommand{\HH}{{\cal H}}

\newcommand{\cd}{\cdot}
\newcommand{\al}{\alpha}

\newcommand{\de}{\delta}
\newcommand{\De}{\Delta}
\newcommand{\ep}{\epsilon}
\newcommand{\ga}{\gamma}

\newcommand{\la}{\lambda}
\newcommand{\Om}{\Omega}
\newcommand{\om}{\omega}
\newcommand{\si}{\sigma}

\newcommand{\xfin}{x_{\rm fin}}
\newcommand{\xin}{x_{\rm in}}
\newcommand{\xt}{\De x}
\newcommand{\Ci}{{\rm Ci}}
\newcommand{\Si}{{\rm Si}}

\newcommand{\tfin}{t_{\rm fin}}
\newcommand{\tin}{t_{\rm in}}
\newcommand{\omt}{\frac{\Omega_{T*}}{\Omega_{\rm rad*}}}
\newcommand{\II}{\mathcal{I}} 
\newcommand{\Det}{\Delta t}
\newcommand{\oms}{\frac{\Omega_{S*}}{\Omega_{\rm rad*}}}

\begin{document}

\setcounter{page}{0}
\thispagestyle{empty}
%\pagestyle{empty}

%%%%%%%%%%%%%%%%%%%%%%%%%%%%%%%%%%%%%%%%%%%%%%%%%%%%%%%%%%%%%%%%%%%%%%%%%%%%%%%
\begin{flushright}
%astro-ph/yymmnnn
CERN-PH-TH/2009-123\\
SACLAY-T09/121
\end{flushright}

\vskip 8pt

\begin{center}
{\bf \Large {
The stochastic gravitational wave background  \\
\vskip 2pt
from turbulence and magnetic fields \\
\vskip 7pt
generated by a first-order phase transition
}}
\end{center}

\vskip 10pt

\begin{center}
{\large Chiara Caprini$^{a}$, Ruth Durrer$^{b}$ and G\'eraldine
  Servant $^{a,c}$ }
\end{center}

\vskip 20pt

\begin{center}

\centerline{$^{a}${\it CEA, IPhT \& CNRS, URA 2306, F-91191 Gif-sur-Yvette, France}}
\vskip 3pt
\centerline{$^{b}$ {\it D\'{e}partement de Physique Th\'{e}orique, Universit\'{e} de Gen\`{e}ve,}}
\centerline{\it  24 Quai E. Ansermet, CH-1211 Gen\`{e}ve, Switzerland}
\vskip 3pt
\centerline{$^{c}${\it CERN Physics Department, Theory Division, CH-1211 Geneva 23, Switzerland}}
\vskip .3cm
\centerline{\tt  chiara.caprini@cea.fr, ruth.durrer@unige.ch, geraldine.servant@cern.ch}
\end{center}

\vskip 13pt

\begin{abstract}
\vskip 3pt
\noindent

We analytically derive the spectrum of gravitational waves due to 
magneto-hydrody-\ namical turbulence generated by bubble collisions in a first-order phase transition. In contrast to previous studies, we take into account the fact that turbulence and magnetic fields act as sources of gravitational waves for many Hubble times after the phase transition is completed.  This modifies the gravitational wave spectrum at large scales. We also model the initial stirring phase preceding the Kolmogorov cascade, while earlier works assume that the Kolmogorov spectrum sets in instantaneously. The continuity in time of the source is relevant for a correct determination of the peak position of the gravitational wave spectrum. We discuss how the results depend on assumptions about the unequal-time correlation of the source and motivate a realistic choice for it.
Our treatment gives a similar peak frequency as previous analyses but the 
amplitude of the signal is reduced due to the use of a more realistic power spectrum for the magneto-hydrodynamical turbulence. For a strongly first-order
electroweak phase transition, the signal is observable with the space interferometer LISA.

\end{abstract}

\newpage

\tableofcontents

\vskip 13pt

%%%%%%%%%%%%%%%%%%%%%%%%%%%%%%%%%%%%%%%%%%%%%%%%%%%%%%%%%%%%%%%%%%%%%%%%%%%%%

\section{Introduction}
Cosmological observations are often a search for `relics' of the early universe.
Relics allow us to infer the physics at a time when the Universe was much 
hotter and much denser than today. A famous example of this is the 
cosmic microwave background (CMB)~\cite{mybook}, which literally represents a 
photograph of the Universe at the time when CMB photons decoupled. Another very
promising relic which has not yet been observed is a gravitational 
wave (GW) background. Since GWs interact so little with matter
and radiation, they propagate freely immediately after generation and 
therefore allow us a direct observation of the Universe at the time of their 
production. GW backgrounds have been proposed from 
inflation~\cite{GWinf}, from braneworlds~\cite{Randall:2006py}, from 
topological defects~\cite{GWtop}, from 
reheating after inflation~\cite{GWreh} and from first order phase 
transitions~\cite{Witten:1984rs,Hogan}.
In this last case, which
is the topic of this paper, at least three 
different sources of GWs have been identified: the collisions of broken phase bubbles ~\cite{Kos1,Kos2,Kos3,Kamion,Caprini:2007xq,Huber:2008hg,thomas}, 
fluid turbulence~\cite{turb1,turb2,CDturb,Tinaturb,Megevand:2008mg,Kahniashvili:2008pe}
and magnetic fields~\cite{GW1mag,mag2,Kahniashvili:2009mf,Caprini:2009pr}. In this paper we concentrate on the
latter two aspects which cannot be truly separated since the magnetic fields 
are processed and amplified by the turbulent fluid flow.

The GW signal from a  first-order 
phase transition has a characteristic frequency of the order
$\om = c\, k_* a_*/a_0 =\ep^{-1}H_*  a_*/a_0$ where $\ep =L_*H_*/c$
is the size of the largest bubbles in units of the horizon size at the 
transition, $1/H_*$. Its value depends on the particle physics model but for a strong first-order phase transition, we will set $\ep\sim 0.01$. 
Quite remarkably, for a potentially first-order electroweak (EW) phase transition at $T_*\sim 100$ GeV,
this frequency is around a MilliHertz  which is the frequency of best 
sensitivity of the planned GW  satellite LISA 
(Laser Interferometer Space Antenna)~\cite{LISA}, meaning that LISA is potentially a window on EW and TeV scale particle physics~\cite{general1,general2,general3}.

The main difference of our approach with earlier works is that we consider a long-lasting, magneto-hydrodynamic (MHD) turbulent source. The collision of  bubbles of the broken phase causes an injection of energy in the primordial fluid. Since the kinetic Reynolds number of the fluid is huge, for instance $\sim 10^{13}$ at the EW epoch (\cf section \ref{sec:howlong}), turbulent motion sets in rapidly in the fluid. The magnetic Reynolds number being also  very large, this leads to the amplification of magnetic fields generated during the phase transition \cite{mag1}, and MHD turbulence develops.
Once the phase transition is over the source of energy injection stops. However, MHD turbulence does not cease immediately but decays like a power law. In previous studies \cite{turb1,turb2,CDturb,Tinaturb,Kahniashvili:2008pe}, the free decay of the turbulent velocity power spectrum has been ignored: it was assumed that turbulence was active only during the completion of the phase transition. However, since the value of the Reynolds number is very high, the dissipation is not sudden and the fluid can remain turbulent during many Hubble times.  

The power law decay in time of MHD turbulence is well established theoretically~\cite{Davidson,MHDturb,olesen,shiro,son,Campanelli:2007tc}, experimentally~\cite{exp} and by numerical simulations~\cite{biskamp,Christensson:2000sp,Banerjee:2004df}. However, there is no consensus on the actual value of the power law exponent: Kolmogorov theory predicts a faster decay than what observed in general in numerical simulations. On the other hand, both analytical analyses and numerical simulations do agree on one point: that, in the absence of helicity, the large scale part of the MHD turbulent spectrum is constant in time. Without inverse cascade the energy is dissipated on the very small scales, the correlation length grows in time, but wavenumbers much smaller than the inverse correlation length are not affected by the evolution. This is important for MHD turbulence in the early universe. The radiation dominated universe is characterised by a finite causal horizon, beyond which the turbulent motions cannot be causally connected. As previously demonstrated \cite{DC03,CDturb}, the existence of this causal horizon in a cosmological setting implies that the real-space correlation function of the stochastic velocity field has compact support. The fact that the real-space correlation function necessarily vanishes at large scales, together with the property of divergence freeness satisfied by the incompressible turbulent flow and by the magnetic field, entails the formation of a Batchelor spectrum for the turbulent velocity field and the magnetic field \cite{CDturb}. Namely, taking for  example the velocity field, the large scale part of the spectrum grows as $P_v(k\rightarrow 0)\propto I\,k^2$, where $I\sim \vev{v^2} L^5$ is the Loitsyansky's integral, $\vev{v^2}$ being the typical velocity of the largest eddies and $L$ their size (\emph{i.e.} the largest scale on which turbulence develops, the correlation scale corresponding to the bubble diameter in our context). Given this form of the large scale part of the power spectrum, if it has to be constant as predicted by the theory and observed in numerical simulations, then necessarily $I$ must be constant in time (which is also required by the Navier-Stokes equation)~\cite{Loi,LLt}. 

In the following, we will assume that the kinetic energy $\vev{v^2}$ and the correlation length $L$ evolve in such a way, as to maintain the product $I\sim \vev{v^2} L^5$ constant. We define the power law exponent $\gamma$ such that $L\sim t^\gamma$ and $\vev{v^2}\sim t^{-5\gamma}$. For generality, the exponent $\gamma$ is kept unspecified in the analytical formulae, but for the numerical results we substitute the value $\gamma=-2/7$. According to Kolmogorov theory, in fact, the constancy of $I$ together with the energy decay equation $d \vev{v^2}  / d t \sim - \vev{v^3}  /L$ lead to the Kolmogorov decay laws: the decay of the kinetic energy with time as $\vev{v^2}\sim t^{-10/7}$ and the growth of the correlation scale as $L\sim t^{2/7}$ (see for example \cite{Davidson}). As explained in section \ref{sec:magnetic}, in the following we also assume equipartition between the turbulent and magnetic energy densities $\vev{v^2}\sim\vev{b^2}$: consequently, we assume the same decay law also for the magnetic field, \ie $\vev{b^2} L^5={\rm constant}$. 

The value $\gamma=2/7$ that we use in the numerical estimates has also been derived on the basis of detailed theoretical arguments, as for example in \cite{Campanelli:2007tc}. On the other hand, as previously mentioned, numerical simulations observe a slower decay for the kinetic and magnetic energies, close to $\vev{v^2}\sim \vev{b^2}\propto t^{-1}$ \cite{biskamp,Christensson:2000sp,Banerjee:2004df}. However, it is not clear whether numerical simulations can efficiently model the conditions of MHD turbulence in the early universe, which is characterised by the presence of a causal horizon and develops at extremely high Reynolds number, of the order of $10^{13}$ (\cf the discussion at the end of section \ref{sec:freedecay}). Therefore, in our analysis we have chosen to follow the theoretical picture of Ref.~\cite{Campanelli:2007tc}, which in addition leads to a conservative estimate of the production of GWs: in fact, MHD turbulence which decays slower would be active as a source of GWs for a longer time. 

In the following we assume this model of free decay for the turbulence, but to this `absolute' time behaviour we also add the exponential de-correlation proposed in \cite{Kraichnan},  to express the time de-correlation of the velocity field on a given scale as time goes by. We take the characteristic de-correlation frequency on a given scale to be the eddy turnover time at that scale, and assume a Gaussian functional form to express the dependence of the power spectrum on time difference. 

Another new point of our analysis is that the sources of GWs, turbulent kinetic energy and magnetic field, are continuous in time. The importance of having a continuous source and its consequences on the position of the peak of the GW spectrum have been analyzed in \cite{thomas} for a short lasting source: the collision of bubbles. Here, we analyze also the long lasting, MHD turbulent source. In order to do so, we need to modify the Kolmogorov decay laws, and insert an initial phase in which the proper turbulent cascade has not yet begun: during this phase, we assume that the kinetic energy starts from zero and grows linearly in time, up to when stirring is over and the free decay of turbulence starts. The linear increase has been observed in MHD simulations \cite{Cho:2008ah}, and is also satisfied in simulations of the bubble collision source (\cf \cite{Huber:2008hg,thomas}). We assume that the evolution law of the stirring scale $L$ remains equal to the Kolmogorov decay law also during the initial phase.

Contrary to previous analyses, we also model the MHD turbulence spectrum using a formula which smoothly interpolates between the large scale behavior, determined by causality, and the small scale one, given by the MHD cascade (see also \cite{Caprini:2009pr}). This model gives a more realistic estimate of the amplitude of the MHD spectrum at the peak, which in turns determines the final amplitude of the GW spectrum. With this improved MHD spectrum, the GW peak amplitude is more than one order of magnitude smaller than previous estimates.

The paper is organized as follows. In the next section we discuss a toy 
model for the source to illustrate the difference between a 
short-lasting and a long-lasting source of GWs. In Section~\ref{s:turb}
we discuss the properties of the MHD turbulence, and we determine the
anisotropic stress power spectrum from this source in 
Section~\ref{sec:anisotropic}. We  then define the
time after which we may neglect the GW source and determine  the final GW 
spectrum in Section~\ref{s:GWspec}. We discuss our results 
in Section~\ref{s:dis} and conclude in Section~\ref{s:con}. A discussion of the fluid viscosity and some 
technicalities are given in appendices for completeness. 

{\bf Notation:} Unless otherwise stated, we use comoving variables: $t$ 
denotes conformal time, the energy injection scale $L$,  the 
Kolmogoroff microscale $\la$ (the endpoint of the Kolmogoroff spectrum), 
and $k$ are respectively comoving distances and wavenumber. The index 
${}_*$ indicates the time of the phase transition, while the index ${}_0$ 
indicates today. We normalize the scale factor $a(t_0)=1$. $\mathcal{H}$ 
denotes the conformal Hubble parameter, $H_0=h_0100$km/s/Mpc is the Hubble 
parameter today, and the critical energy density is $\rho_c=\rho_c(t_0)$. 
The radiation energy density parameter today is 
$h_0^2\Om_{\rm rad, 0}=4.2\times 10^{-5}$.
This value includes three types of neutrinos. As we shall see, this is
the relevant quantity since neutrinos (with standard masses) are still 
relativistic at matter-radiation equality. 
%During the radiation dominated universe relevant in this discussion $\HH=1/t$. 

\section{A stochastic gravitational wave background of cosmological origin}

We consider a Friedmann universe with flat spatial sections. The tensor metric perturbations are defined by
\be
ds^2=a^2(t)[-dt^2+(\de_{ij}+2h_{ij}) dx^idx^j]\,.
\ee
In this work we want to determine the GW energy density power spectrum
given by (see e.g.~\cite{GW1mag})
\be
\frac{d \Omega_{GW}}{d \log k}=\frac{k^3|\dot{h}|^2}{2(2\pi)^3G\rho_ca^2}=
   \frac{k^5|{h'}|^2}{2(2\pi)^3G\rho_ca^2}\, , \qquad \rho_c=\rho_c(t_0)\,,
\ee
where $\dot{}=\frac{d}{d t}$ and $'=\frac{d}{d x}$, $x=k t$, and the 
GW energy power spectrum is defined as
\be
\vev{\dot{h}_{ij}(\bk,t)\dot{h}_{ij}^*(\bq,t)}= (2\pi)^3\de(\bk-\bq)
   |\dot{h}(k,t)|^2 \,.
\ee
Here $\vev{\cdots}$ is an ensemble average over the stochastic process which 
generates the GWs. The Dirac delta function $\de(\bk-\bq)$ is a
consequence of statistical homogeneity.

Once the source has decayed and the wavelength under
 consideration is inside the horizon, the GW energy
 density simply scales like $a^{-4}$. Hence the GW energy spectrum scaled to
 today becomes
\be
\left.\frac{d\Omega_{GW}}{d \log k}\right|_0  =
 \left.\frac{d\Omega_{GW}}{d \log k}\right|_{t} a(t)^4 =
 \frac{k^5 a^2 }{2(2\pi)^3G\rho_c}|{h'(x)}|^2\,, \qquad x\gg 1.
\label{omegatoday}
\ee

To evaluate the GWs emitted by turbulent motion in 
the primordial fluid and by a  magnetic field we need to determine the 
tensor-type anisotropic stresses of these sources. They source the evolution
equation for the GW perturbations,
\be
\ddot{h}_{ij}+2\mathcal{H} \dot{h}_{ij}+k^2h_{ij}=8\pi G a^2 T^{(TT)}_{ij}(k,t)
 \,.
\label{waveeq}
\ee
In this section we 
consider in all generality a relativistic source, and we solve the 
wave equation in two cases: a long lasting source (\ie many Hubble times), 
and a short lasting one (\ie significantly less than one Hubble time). We 
introduce the transverse traceless tensor part of the energy momentum tensor 
of the source as
\be
T^{(TT)}_{ij}(k,t)=(\rho+p)\tilde{\Pi}_{ij}(k,t) \qquad \mbox{ so that } \qquad
8\pi G a^2 T^{(TT)}_{ij}(k,t) = 4\mathcal{H}^2\tilde{\Pi}_{ij}(k,t)\,,
\label{TTT}
\ee
where we denote the 
dimensionless energy momentum tensor with a tilde:
$\tilde \Pi_{ij}(\bk,t)=(P_{il}P_{jm}-1/2P_{ij}P_{lm})\tilde T_{lm}(\bk,t)$. The projection
tensor $P_{il}P_{jm}-1/2P_{ij}P_{lm}$, with $P_{ij}=\delta_{ij}-\hat{k}_i\hat{k}_j$,  
projects onto the transverse traceless part of the stress tensor. 
$\tilde{\Pi}$ includes any time dependence other than the basic 
radiation-like evolution. We assume that the source is active only during the 
radiation-dominated era, where $p=\rho/3$. During adiabatic expansion 
$g(Ta)^3=$ constant so that
\be
\rho(t)=\frac{\rho_{\rm rad,0}}{a^4(t)}
  \left(\frac{g_0}{g(t)}\right)^{1/3} \ \ \
 \mbox{and} \ \ \  \ \fbox{$
a(t) \approx  H_0 \ \Omega_{\rm rad,0}^{1/2}
   \left(\frac{g_0}{g(t)}\right)^{1/6} t$}
\label{rhoa} \ee
where $g(t)$ is the number of relativistic degrees of freedom at time $t$.

\subsection{Long-lasting source}

Let us first concentrate on the more general case of a long lasting source. 
To solve Eq.~(\ref{waveeq}) we set $\mathcal{H}=1/t$, neglecting changes in 
the number of effective relativistic degrees of freedom. 
In terms of the dimensionless variable $x=k t$ Eq.~(\ref{waveeq}) then becomes
\be
 h''_{ij}+2 \frac{h'_{ij}}{x}+h_{ij}=\frac{4}{x^2}\tilde{\Pi}_{ij} \, .
\ee
We consider a source that is active from time $\tin$ to time $\tfin$, 
which in the long lasting case can span a period of many Hubble 
times. For $t>\tfin$, we match the solution of the above equation to 
the homogeneous solution, $\tilde{\Pi}_{ij}=0$. Assuming further that we are 
only interested in modes well inside the horizon today, $x\gg1$, the 
resulting GW energy power spectrum becomes
\be\label{e:h'spec}
\left|h'(k,x>x_{\rm fin})\right|^2=\frac{8}{x^2} \int_{x_{\rm in}}^{x_{\rm fin}} 
 \frac{dx_1}{x_1} \int_{x_{\rm in}}^{x_{\rm fin}} \frac{dx_2}{x_2}
\cos(x_2-x_1) \tilde{\Pi}(k,x_1,x_2) \, \qquad x\gg 1\,,
\ee
 $x_1=kt_1$, $x_2=kt_2$, and $\tilde{\Pi}(k,x_1,x_2)$ denotes the unequal time correlator of the source, 
\be
\langle \tilde{\Pi}_{ij}({\bf{k}},t_1)\tilde{\Pi}_{ij}^*({\bf{q}},t_2) 
 \rangle= (2\pi)^3 \delta({\bf{k}}-{\bf{q}}) \tilde{\Pi}(k,kt_1,kt_2) \, .
\label{Pispectrum}
\ee
With Eq.~(\ref{omegatoday}), the power spectrum of the GW energy 
density parameter for a long-lasting source which is active between 
$t_{\rm in}$ and $t_{\rm fin}$ in the radiation era is then given by
\be
\label{longlastingcase}
\left.\frac{d\Omega_{GW}}{d \log k}\right|_0 =
 \frac{4 \ \Omega_{\rm rad,0} }{3 \pi^2} 
 \left(\frac{g_0}{g_{\rm fin}}\right)^{1/3} \ k^3 \  
\int_{x_{\rm in}}^{x_{\rm fin}} \frac{dx_1}{x_1} \int_{x_{\rm in}}^{x_{\rm fin}} 
\frac{dx_2}{x_2}\cos(x_2-x_1) \tilde{\Pi}(k,x_1,x_2) \,.
\ee
This result is completely general for modes well inside the horizon today; it reduces the computation of the GW 
spectrum to the determination of the unequal-time correlator of the 
tensor-type anisotropic stress, $\tilde{\Pi}(k,x_1,x_2)$.

\subsection{Short-lasting source}
\label{sec:short_lasting}

If the source is active for a short interval of time, essentially 
only during the phase transition, one can neglect the 
expansion of the universe during the time of action of the source, and match 
the solution so obtained with the one of the homogeneous equation in which 
expansion is taken into account. For this kind of source, we set  
$t_{\rm fin}=\tin+\De t$, with $\De t/\tin \ll 1$. Solving the wave equation 
(\ref{waveeq}) without expansion term, amounts to neglect 
the time-dependence of the factors $1/x_1$ and $1/x_2$ in Eq.~(\ref{e:h'spec}) 
or (\ref{longlastingcase}) during the active period, so that
\be
\left|h'(k,x>x_{\rm fin})\right|^2=\frac{8}{x_{\rm in}^2\,x^2}
\int_{x_{\rm in}}^{x_{\rm fin}} {dx_1} \int_{x_{\rm in}}^{x_{\rm fin}} {dx_2}
\cos(x_2-x_1) \tilde{\Pi}(k,x_1,x_2)\,.
\ee
Also this solution applies for modes inside the horizon today. 
The energy spectrum now becomes
\be
\left.\frac{d\Omega_{GW}}{d \log k}\right|_0 =
\frac{4 \ \Omega_{\rm rad,0} }{3 \pi^2} 
 \left(\frac{g_0}{g_{\rm fin}}\right)^{1/3} \ {\cal H}_{\rm in}^2 \,k \  
\int_{x_{\rm in}}^{x_{\rm fin}} {dx_1} \int_{x_{\rm in}}^{x_{\rm fin}} {dx_2}
\cos(x_2-x_1) \tilde{\Pi}(k,x_1,x_2)\,. \label{shortla}
\ee
Obviously, the general long-lasting case reduces to this result if 
$(t_{\rm fin}-t_{\rm in})/t_{\rm in} \ll 1$.
\vspace*{1cm}
\noindent
Summarizing:\\

\hspace*{-0.6cm}\fbox{\vbox{
\begin{eqnarray}
\left. \frac{d\Omega_{GW}h_0^2}{d\log k}\right|_{0}=\mathcal{A} 
\left\{\begin{array}{l}
\vspace*{0.3cm}
\left(\frac{g_0}{g_{\rm fin}}\right)^{\frac{1}{3}} \ k^3
\int_{x_{\rm in}}^{x_{\rm fin}} \frac{dx_1}{x_1}\int_{x_{\rm in}}^{x_{\rm fin}} 
\frac{dx_2}{x_2}\, \cos(x_2-x_1)\,\tilde{\Pi}(k,x_1,x_2) \\
\hspace*{3.5cm}\mbox{\small long-lasting source
  (\eg MHD turbulence)}\\       \\
\vspace*{0.3cm}
\left(\frac{g_0}{g_*}\right)^{\frac{1}{3}}
\mathcal{H}_{\rm in}^2\,k\int_{x_{\rm in}}^{x_{\rm fin}} dx_1 \int_{x_{\rm in}}^{x_{\rm fin}}  dx_2\, \cos(x_2-x_1)\,\tilde{\Pi}(k,x_1,x_2) \\
\hspace*{2.5cm}\mbox{\small  short-lasting source
(\eg bubble~collisions)} 
\end{array}\right.
\label{definition}
\end{eqnarray}
~~~~with $x_1=k t_1$, $x_2=k t_2$, $\xin=k\tin$, $\xfin=k\tfin$ and 
$\mathcal{A}=\frac{4}{3\pi^2}\Omega_{\rm rad,0}h_0^2\,.$ }}

\subsection{Solutions for a simple source}
\label{sec:constant}

In this section we analyze the difference between the GW spectrum generated by 
a short lasting and a long lasting  source in a simple example which can be 
treated analytically. We find analytical solutions for the GW energy density 
spectrum (\ref{definition}). The general behaviour of the solutions in this simple case is illuminating, as
it is similar to the case of the evolving source that we will treat in 
the rest of the paper (MHD turbulence). 

We consider a tensor source $\tilde{\Pi}$ with an equal time power spectrum 
which depends on time solely via a function modeling the turning on and off of 
the source: $\tilde{\Pi}(k,t,t)\equiv \tilde{\Pi}(k)f^2(t)$. 
For the  equal time power spectrum of the tensor source, we take a form which is motivated by 
turbulence and magnetic fields, see Sec.~\ref{sec:anisotropic}: 
\be
\tilde{\Pi}(K,t,t)=\left(\frac{\Omega_{S}}{\Omega_{\rm rad}}\right)^2 L^3 
   \mathcal{S}(K)f^2(t)\,,
\ee
where $\Omega_S$ denotes 
the (radiation like) energy density of the source normalised to the critical energy density today (so that the ratio 
$\Omega_{S}/\Omega_{\rm rad}$ is time-independent), $K=Lk/2\pi$ is a dimensionless wavenumber,  $L$ is a characteristic 
scale of the problem, and $\mathcal{S}(K)$ models the scale dependence of the 
source. The continuous  function $f(t)$ vanishes at both, $\tin$ and $\tfin$ and describes the 
switching on and off of the source. The source is active during the time interval $\tfin-\tin$, which can be 
long or short compared to the initial Hubble time, $\mathcal{H}_{\rm in}^{-1}=\tin$. 

\begin{figure}[htb!]
\begin{center}
\includegraphics[width=6.1cm]{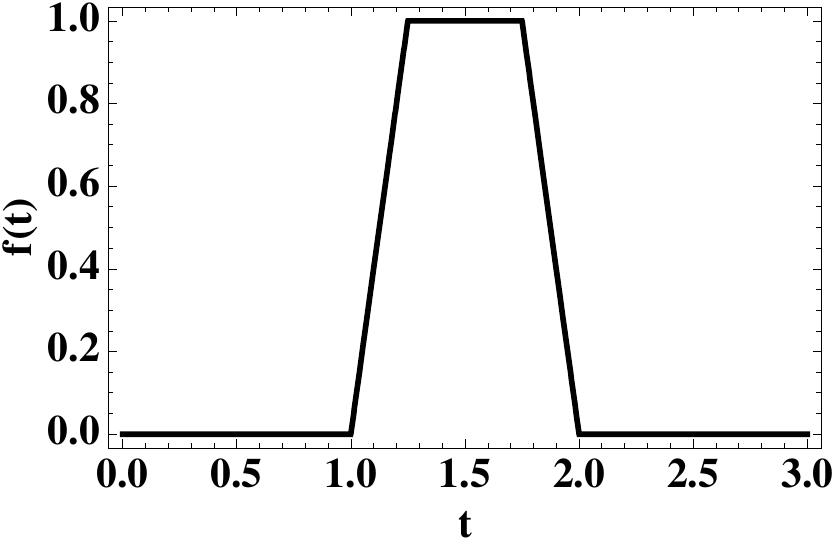}
\caption{\label{f_of_eta} \small The function modeling the time dependence of the source used in Section \ref{sec:constant}, Eq.~(\ref{e:ftau}).}
\end{center}
\end{figure}

As shown in Ref.~\cite{thomas}, the time continuity in switching the source on and off can be relevant for the GW spectrum. Inspired by results from numerical simulations of bubble nucleation during a 
first order phase transition \cite{Kos3,Huber:2008hg}, we choose the function $f(t)$ to be continuous but not differentiable at the initial 
and final times. The 
effect of this choice on GW spectra for short duration sources is discussed 
in Ref.~\cite{thomas}. Here we shall also study its effect on sources of 
long duration.
We set (see Fig.~\ref{f_of_eta})
\be\label{e:ftau}
f(t)=\left\{ \begin{array}{ll}
 0 &\mbox{if } t\le \tin \\
 \frac{2(t-\tin)}{\Det} &\mbox{if } \tin\le t\le \tin + 
            \Det/2 \\
 1 &\mbox{if } \tin+ \Det/2 \le t\le \tfin - \Det/2 \\
  \frac{2(\tfin-t)}{\Det} &\mbox{if } \tfin-\Det/2 
 \let\le \tfin \\
 0 &\mbox{if } t\ge \tfin \,.
\end{array}\right.
\ee
For a short lasting source, $\Det$ is also the duration of the source 
(\cf Sec.~\ref{sec:short_lasting} and Ref. \cite{Caprini:2007xq}), while a long 
lasting source is typically  active for a much longer period of time, and $\Det$ is the characteristic time of turning on and off. As we show below, the effect of introducing time continuity on the resulting GW spectrum is relevant only in the `coherent' case (see Fig.~\ref{fig:discontinuous}).  
In this case, the power at small scales is less than for a discontinuous source. 

To compute the solution according to Eq.~(\ref{definition}), we need the 
source power spectrum at unequal times $\tilde{\Pi}(k,t_1,t_2)$. In this 
section, we consider two different possibilities for the unequal time 
correlators, the incoherent and the totally coherent approximations, which we 
discussed also in the case of GWs generated by bubble 
collisions~\cite{Caprini:2007xq}. In Sec.~\ref{s:turb} we will compare these 
approximations with the `top hat' ansatz~\cite{Caprini:2007xq}, which turns 
out to be more realistic for the case of MHD turbulence. 

\vspace*{0.5cm}
\noindent
$\bullet$~{\bf  Incoherent approximation}: the source is correlated only 
for $t_1\simeq t_2$
\be
\tilde{\Pi}(K,t_1,t_2)=\tilde{\Pi}(K,t_1,t_1)\delta(t_1-t_2)\Det \, f^2(t_1)\,.
\label{inco1}
\ee
Here $\Det$ plays also the role of a very short characteristic time, 
over which the source remains `coherent'. With this ansatz the time 
integration in (\ref{definition}) is simple, and we obtain the following 
result for the GW energy spectrum:
\begin{eqnarray}
\left. \frac{d\Omega_{GW}h_0^2}{d\log k}\right|_{0}&=&
\frac{4\Om_{\rm rad,0}h_0^2}{3\pi^2}  
\left(\frac{\Omega_{S}}{\Omega_{\rm rad}}\right)^2 K^3 \mathcal{S}(K)
  F(\tin,\tfin,\Det) 
\end{eqnarray}
where
\begin{eqnarray}
 F(\tin,\tfin,\Det) &\simeq& 
\left\{\begin{array}{l}
\vspace*{0.3cm}
 \big(\frac{g_0}{g_{\rm fin}}\big)^{\frac{1}{3}}\, \frac{\Det}{\tin}  \qquad    
   \qquad \quad\mbox{long-lasting,} \vspace*{0.3cm}\\
 \big(\frac{g_0}{g_*}\big)^{\frac{1}{3}}\,
 \frac{(2\pi)^2}{3}\,\big(\frac{\Det}{\tin}\big)^2 ~ \qquad\mbox{short-lasting.}
\end{array}\right.
\label{constantshort}
\end{eqnarray}
The result for the long-lasting case is expanded using $\Det\ll \tin,\tfin$. 
The full expression is given in Appendix~\ref{Appen:analyt}, where we collect 
all analytical expressions for the convenience of the reader, see 
Eq.~(\ref{ap:constantshort}). 

In the incoherent approximation, the function $F(\tin,\tfin,\Det)$ resulting 
from the convolution of the Green function with the source,
Eq.~(\ref{definition}), does not depend on wave-number. The GW power spectrum 
is therefore simply the one of the source $S(K)$, multiplied by the phase-space volume 
$K^3$, both for long lasting and for short lasting sources. The peak of the GW 
spectrum then coincides with the one of the source spectrum.
The time integration results in the ratio between the brief 
coherence time $\Det$ and the initial horizon time $\tin$. In the long 
lasting case, this factor might be close to one while in the short lasting case
it is always much smaller. In addition, the GW amplitude of the
short lasting case is suppressed by one more factor $\Det/\tin$ with respect to the long lasting one. 

\vspace*{0.5cm}
\noindent
$\bullet$~{\bf  Coherent approximation}: the source is perfectly correlated at
all times $t_1$ and $t_2$
\be
\Pi(K,t_1,t_2)=\sqrt{\Pi(K,t_1,t_1)}\sqrt{\Pi(K,t_2,t_2)}\,.
\label{co1}
\ee
In this case as well, the time integration in Eq.~(\ref{definition}) can be 
performed explicitely,
and we obtain the GW energy density power spectrum (we remind that 
$\xin=kt_{\rm in}$, $\xfin=kt_{\rm fin}$, and $\De x=\xfin-\xin$)
\begin{eqnarray}
\left. \frac{d\Omega_{GW}h_0^2}{d\log k}\right|_{0}&=&
\frac{4\Om_{\rm rad,0}h_0^2}{3\pi^2}  
\left(\frac{\Omega_{S}}{\Omega_{\rm rad}}\right)^2 K^3 \mathcal{S}(K)
   F(\xin,\xfin,\De x) \label{factorcoherent}
\end{eqnarray}
where
\begin{eqnarray}\label{constantco}
  F(\xin,\xfin,\De x) &\simeq&
\left\{\begin{array}{ll}
\vspace*{0.3cm}
\big(\frac{g_0}{g_{\rm fin}}\big)^{\frac{1}{3}}\,\left[
 \left(\Ci(\xfin) - \Ci(\xin)\right)^2 +  \left(\Si(\xfin) - 
    \Si(\xin)\right)^2\right] &\mbox{long-lasting}\\
\big(\frac{g_0}{g_*}\big)^{\frac{1}{3}}\,
\frac{64(2\pi)^2}{\xin^2}\,
    \frac{\sin^4((\xfin-\xin)/4)}{(\xfin-\xin)^2}  & \mbox{short-lasting.}
\end{array}\right.
\end{eqnarray}
In the long lasting case we have again expanded to lowest order in $\xt/\xin$ 
and $\xt/\xfin$. $\Ci$ and $\Si$ denote the integral cosine and sine 
functions~\cite{AS}. The full expression is given in 
Appendix~\ref{Appen:analyt}.

\begin{figure}[htb!] 
\begin{center}
\includegraphics[width=10cm]{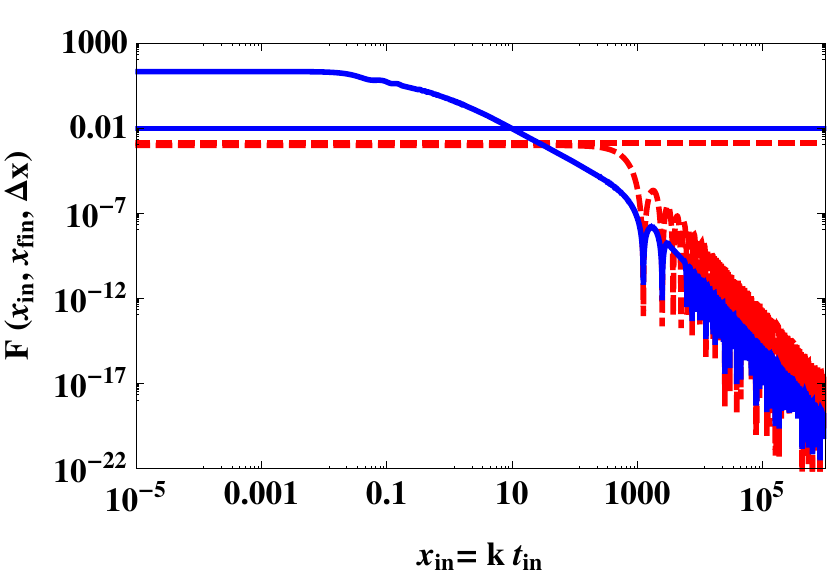}
\caption{\label{fig:factorcoherent} \small Comparison between a short and long-lasting source. We plot the function  $F(\xin,\xfin,\De x)$ 
defined in Eq.~(\ref{constantshort}) (incoherent case) and Eq.~(\ref{constantco}) (coherent case) as a function of $\xin=k\tin$. Blue, solid: long lasting coherent and incoherent 
cases with $\tfin/\tin=100$ and $\De t=0.01\, \tin$. Red, dashed: short lasting coherent and incoherent cases 
with $\tfin/\tin=1.01$, $\De t=0.01\, \tin$. The horizontal lines correspond to the incoherent case.}
\end{center}
\end{figure}

\begin{figure}[htb!] 
\begin{center}
\includegraphics[width=10cm]{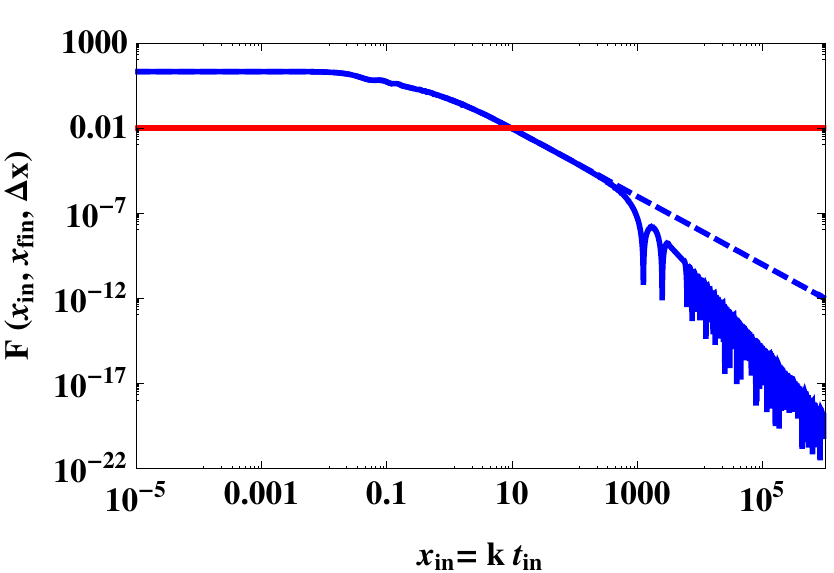}
\caption{\label{fig:discontinuous} \small For a long-lasting source, comparison between the continuous (solid) and discontinuous (dashed) cases. We plot the function  $F(\xin,\xfin,\De x)$ as a function of $\xin=k\tin$. The incoherent case is not affected by continuity as shown by the red horizontal line  (the solid and dashed lines are superimposed). In the coherent case shown in blue, the slope changes in the continuous case for frequencies $k>\Det^{-1}$. The values of the parameters are the same as in
 Fig.~\protect\ref{fig:factorcoherent}.}
\end{center}
\end{figure}

Contrary to the incoherent case, here the function $F(\xin,\xfin,\De x)$ 
depends on wave-number, and the resulting GW spectrum is therefore modified 
with respect to the one of the source. We plot $F(\xin,\xfin,\De x)$ in Fig.~\ref{fig:factorcoherent}. In both the long and short lasting 
case, $F(\xin,\xfin,\De x)$ tends to a $k$-independent value for wave-numbers 
such that $\xin\ll 1$. The constant is $\log^2(\xfin/\xin)$ in the long 
lasting case, and $\pi^2(\De x/\xin)^2$ in the short lasting one. In the short
lasting case $F$ remains constant up to $\De x\simeq 1$ where it starts 
oscillating and decaying like $1/(\xin\De x)^2$. The long lasting case instead 
depends also on $\xfin$: when $\xfin$ becomes larger than one, the slope changes from the constant to a mild logarithmic dependence on $k$ as $\log^2\xin$. Then for $\xin\geq 1$ we have a decay like $1/\xin^2$, up to $\De x\gtrsim 1$ where $F$ also starts oscillating and decaying, with a smaller amplitude than the short lasting case, see Fig.~\ref{fig:factorcoherent}. Hence, wavelengths which are larger than the typical switching on time are amplified  by an
additional factor $\min\{(\De x)^{-2},(\xin/\De x)^2\}$ in the long 
lasting case. On the other hand, wavelengths smaller than the the typical switching on time are suppressed in the long lasting case due to the presence of interferences suppressing the signal.

The functions $F(\xin,\xfin,\De x)$ for both the short and long lasting 
incoherent and coherent cases are shown in Fig.~\ref{fig:factorcoherent}.
The GW amplitude in the short lasting case is suppressed for low 
and intermediate values of $k$ with respect to the long lasting one, both in 
the incoherent and in the coherent cases. However, while in the incoherent case this suppression is maintained for all $k$, 
in the coherent one interferences suppress the amplitude of the long lasting case for frequencies larger than the typical 
switching on frequency $\De x \gtrsim 1$.   


Fig.~\ref{fig:discontinuous} shows the effect of introducing continuity in the process of turning on and off  a long lasting source. The conclusions drawn in the analysis of Ref.~\cite{thomas} are not modified by the long duration of the source. The incoherent spectrum is the same for a continuous and a discontinuous source, while in the coherent case there is a difference for frequencies higher than $k\gtrsim 1/\Det$ corresponding to the characteristic time-scale of the source. In the continuous case the slope changes, becoming steeper by a factor $k^{-1}$ if the turning on process has a kink: $f(t)$ is continuous but not differentiable. In the discontinuous case, on the other hand, the change of slope is absent.

Summarizing, two general features can be deduced from this analysis. First, 
the GW energy spectrum at large scales is proportional to the phase space 
volume $K^3$ times the source power spectrum $\mathcal{S}(K)$. In the 
incoherent case no other wavenumber-dependence intervenes; in the coherent 
case, instead, the wavenumber at which this behaviour changes depends on the 
duration of the source, whether it is short or long lasting. Second, the GW 
amplitude in the short lasting case is smaller than the one in the long 
lasting case. In the incoherent case, this is always true; in the coherent 
case, this suppression can be very significant on large scales, however, it 
turns into an amplification for scales smaller than $\Det$. On super-horizon 
scales, the difference in the amplitude between  the short lasting and  
long lasting sources is typically of the order of $\Det/\tin$ in the incoherent 
case and  $(\Det/\tin)^2$ in the coherent one.

Examples of well-motivated long-lasting sources are turbulent fluid flows 
initiated by instabilities generated by bubble collisions. In the following, 
we concentrate on this particular source. We also study the case of magnetic 
fields, which represent another long-lasting source of GW typically expected 
from first order phase transitions~\cite{mag1}.

\section{Turbulent magneto-hydrodynamics as a source of gravitational waves}\label{s:turb}

\subsection{General considerations}
\label{sec:turbuinitial}

Turbulence develops if a fluid with  sufficiently high Reynolds number is 
perturbed. If the fluid is stirred on a characteristic scale $L_p$ (the 
subscript indicates that we use the physical length, not comoving length here),
the Reynolds number of the flow is defined by
\be
\Re(L_p)=\frac{v_{L} \,L_p}{\nu}
\label{Reynold_number}
\ee
where $v_{L}$ is the characteristic velocity on the energy injection scale 
$L_p$, and $\nu$ is the  kinetic viscosity of the fluid. If $\Re(L_p)\gg 1$, 
the stirring develops turbulent motions. In a first order phase transition, 
the source of stirring is bubble collision. Therefore, 
the characteristic scale of the stirring is given initially by the typical 
bubble size towards the end of the phase transition, 
$L_*\sim 2 v_b {\beta}^{-1}$ where $v_b$ is the bubble wall velocity 
and $\beta^{-1}$ the (comoving) duration of the phase transition (see 
for instance Section 4 of \cite{Caprini:2007xq} for a detailed definition). The 
initial energy injection scale $L_*$ is the scale at which the largest turbulent
eddies develop, and corresponds to the peak of the turbulent velocity power spectrum. 

The dynamics of bubble growth at late times, towards the end of the 
phase transition, allows us to determine the order of magnitude of the 
kinetic energy involved in the turbulent flow. The bubble wall can be treated 
as a discontinuity, \ie a combustion front across which energy and momentum 
are conserved \cite{Steinhardt:1981ct}. At the front, the velocity of the fluid in the rest 
frame of the bubble center is given by $v_f=(v_1-v_2)/(1-v_1v_2)$, where $v_1$
 and $v_2$ are respectively the incoming and outgoing speed of the fluid in 
the rest frame of the front (see  e.g. Ref.~\cite{Caprini:2007xq} for more details).
 We assume that the typical value of the turbulent fluid velocity is 
given by $v_f$. Therefore, the kinetic energy of the turbulent flow is 
\be
\rho_{\rm kin}= (\rho+p) \frac{\langle v^2 \rangle}{2}~~~~~~{\rm with}~~~
                \vev{v^2}\sim v_f^2\,.
\label{rhokin}
\ee
Even though the velocities involved are large in the case of interest, we are using the formalism of non-relativistic MHD turbulence. Since the corresponding values of $\gamma$ are typically of order one, we expect that the error introduced is within the uncertainty of our calculation. Furthermore, 
while the validity of the theory of non-relativistic turbulence may be questionable 
if high speeds are involved \cite{Dettmann}, it was shown in \cite{Cho:2004nn,Micha:2002ey, Micha:2004bv} that the Kolmogorov spectrum is recovered even in the relativistic case\footnote{It is remarkable that recent simulations of quark gluon plasma instabilities in the process of  thermalization in heavy ion collisions show similarities with a  Kolmogorov scaling\cite{Arnold:2005qs,Mueller:2006up,Berges:2008mr}.}.
Still, we impose for the fluid velocity $\vev{v^2}\leq c_s^2$, 
where $c_s=1/\sqrt{3}$ is the sound speed in the relativistic fluid. 
The fluid velocity $v_f$ is completely specified once $v_b$ 
is known and the ratio 
$\al=\rho_{\rm vac}/\rho_{\rm rad,*}$ is fixed: solving for the hydrodynamical equation allows one to relate the fluid velocity to the bubble wall velocity 
 (see for instance~\cite{Thomasetal} for more details). If the phase transition proceeds 
as a detonation (deflagration), then $v_1\equiv v_b$ ($v_2\equiv v_b$). For 
this paper, we choose the fluid velocity $v_f=c_s$  corresponding  to 
$\al=1/3$  and either $v_b\simeq 0.87$ (detonation) or $v_b\simeq c_s$ 
(deflagration) (for $\alpha>1/3$ there is no deflagration solution). When we need to specify $v_b$, in the figures and the numerical values, we always consider the detonation case $v_b=0.87$. It is straight forward to re-scale the results to lower bubble and fluid velocities.

If the fluid is stirred on the scale $L_*$, turbulent motions develop within a time interval of the order of the eddy turnover time $\tau_L$. This is the characteristic time for the cascade to set in. Given the typical value of the turbulent fluid velocity $v_f$, the eddy turnover time on the stirring scale $L_*$ is defined simply as $\tau_L\sim L_*/(2v_f)$. Since the fluid velocity is always smaller than the bubble wall velocity $v_f\leq v_b$, the eddy turnover time is always larger than the duration of the phase transition: $\tau_L\geq \beta^{-1}$. In the following we identify the time interval $\Det$ given in Sec.~\ref{sec:constant} as $\Det=\tau_L$. For short-lasting turbulence, this means that the source lasts for only one eddy turnover time. In the long-lasting case (which is the relevant one as we will see), this means that turbulence is `turned on' in one eddy turnover time $\tau_L$.

In the cosmological context the fluid is ionized and has not only a very high
kinetic Reynolds number (which we evaluate in Section~\ref{sec:howlong}) but 
also a very high magnetic Reynolds number ${\rm R_m}$, defined by
(see Appendix~\ref{Appen:viscosity})
\be\label{e:Rm}
{\rm R_m}(L_p) = \frac{L_p v_L}{\mu}\,, \quad  \mbox{ where } \qquad \mu =\frac{1}{4\pi\si}
\ee
is the magnetic diffusivity and $\si$ denotes the conductivity. High values 
of the magnetic Reynolds number require an MHD treatment of the cosmic plasma. 
Moreover, the magnetic Prandl number  
\be\label{e:Pm}
{\rm P_m} \equiv \frac{{\rm R_m}(L)}{\Re(L)} = \frac{\nu}{\mu} 
\ee 
is much larger than unity, as we calculate in 
Appendix~\ref{Appen:viscosity}. Therefore, the characteristics of the plasma 
in the early universe entail the formation of MHD turbulence. The seed magnetic
 field can be generated by several mechanisms \cite{mag1}, and is then 
amplified by the currents due to the turbulent flow of charged 
particles~\cite{MHDturb,BS}. The magnetic field itself is also a source of GWs.
In the following we make the simplifying but reasonable assumption of 
equipartition: the total kinetic energy in the turbulent motion is equal to the 
magnetic field energy. Note however, that this assumption need not hold for 
each wave number $k$, so that we can allow for different spectra for the 
magnetic field energy and the turbulent kinetic energy at small scales. 

In the remainder of this section we present our model for the power spectra of 
turbulence and magnetic fields, as well as their time evolution. We evaluate 
the Reynold number in the early universe and describe how it evolves with time,
which will help us to determine when MHD turbulence is expected to end.

\subsection{The turbulent velocity power spectrum}
\label{sec:turbulence}

\begin{figure}[htb!]
\begin{center}
\includegraphics[width=9.1cm]{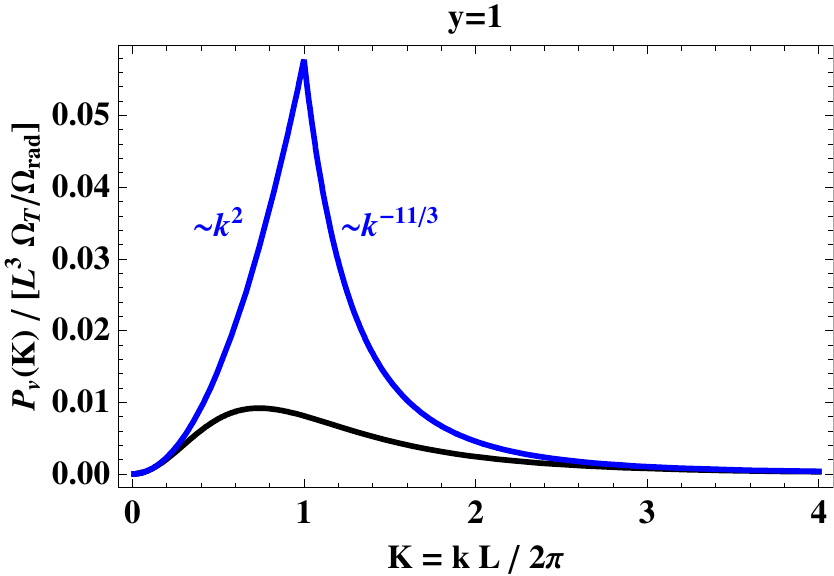}
\caption{\label{comparison} \small Comparison between the turbulent velocity power spectrum obtained by intersecting the $k^2$ behaviour at small scale with the inertial range $k^{-11/3}$ behaviour (blue), as done in the literature, with the Von K\'{a}rm\'{a}n spectrum (black).}
\end{center}
\end{figure}

The power spectrum of the turbulent velocity field at equal times is of the form\footnote{In this paper we neglect the presence of a helical component in the velocity and magnetic field power spectra (\cf Eq.~(\ref{bequaltime})). Non-zero helicity, possibly arising from a macroscopic parity violation in the early universe, affects the decay of MHD turbulence, as demonstrated for example in \cite{biskamp}, and the subsequent generation of GWs. For GW production by primordial helical MHD turbulence we refer to the analysis of Refs.~\cite{Kahniashvili:2008pe,Caprini:2009pr}.}
\be
\vev{v_{i}(\bk,t)v_{j}^*(\bq,t)}=(2\pi)^3 \de(\bk-\bq)\,P_{ij}\,P_v(k,t)\,,
\label{velequaltime}
\ee 
where the projector $P_{ij}=\de_{ij}-\hat{k}_i\hat{k}_j$ comes
from the fact that the turbulent velocity field ${\bf v}(\bx,t)$ is 
divergence free. An ansatz for the unequal time correlator will be given in 
Sec.~\ref{sec:decorrelation}. In previous works, the turbulent velocity 
power spectrum $P_v(k,t)$ was either assumed to be given only by the inertial range 
$k^{-11/3}$ \cite{turb1,turb2,Tinaturb,Kahniashvili:2008pe}, or naively determined by intersecting the $k^2$ behaviour 
at very small scales with the inertial range $k^{-11/3}$ behaviour \cite{CDturb}. 
Both approaches overestimate the peak amplitude of the turbulent source and thus
overestimate the GW amplitude. In the present study, 
we use a more realistic, smooth function to describe the spectrum, proposed 
by Von K\'arm\'an \cite{Karman} (see also page 244 of \cite{Hinze}). It is given by the following
interpolating formula (here in terms of comoving quantities) :
\bea
P_v(K)=\mathcal{C}_v \langle v^2 \rangle \,L^3\,\frac{K^2}{(1+K^2)^{17/6}}\times 
\left\{\begin{array}{ll}
1 & {\rm for}~0\leq K\leq \frac{L}{\lambda} \\
0 & {\rm for}~K\geq \frac{L}{\lambda}\, ,
\label{velocityspectrum}
\end{array}\right.
\eea
where we again use the dimensionless variable 
\be
K=kL/2\pi
\label{BigK}
\ee
and $\lambda$ denotes the Kolmogorov microscale, beyond which turbulent motions are absent and we set the spectrum to zero. The kinetic energy of the turbulent flow is given in Eq.~(\ref{rhokin}): using this definition, we 
rewrite $ \langle v^2 \rangle$ in terms of the ratio of the total kinetic 
energy to the radiation energy density,
\be
\langle v^2 \rangle=\frac{3}{2} 
  \frac{\Omega_{T}}{\Omega_{\rm rad}}\,.
\label{OmTparameter}
\ee
In our numerical estimates, we will use $\langle v^2\rangle=1/3$, thus corresponding to $\Omega_{T}/\Omega_{\rm rad}=2/9$.
In equation (\ref{velocityspectrum}), the constant $\mathcal{C}_v=\frac{55}{108\pi^{3/2}}\,
\frac{\Gamma(5/6)}{\Gamma(1/3)}\approx0.0385$ comes from the normalization of 
   the kinetic energy spectrum, $E(k)=k^2 P_v(k)/(2\pi^2)$,
\be
\frac{\rho_{\rm kin}}{\rho+p}=\frac{\vev{v^2}}{2}=\frac{3}{4}
 \frac{\Omega_T}{\Omega_{\rm rad}}=\int_0^\infty dk E(k)=
        \frac{1}{2\pi^2}\int_0^\infty dk\, k^2 P_v(k) ~.
\label{kinen}
\ee
Expression (\ref{velocityspectrum}) smoothly interpolates between the large 
scale, $K^2$ behaviour, and the inertial range $K^{-11/3}$, which is reached 
for $K\gtrsim 3$. The peak is at $K_{peak}=\sqrt{6/11}\sim 0.74$, which 
corresponds roughly to the energy injection scale at $K=kL/2 \pi =1$. The peak amplitude is smaller by a factor $\sim 6$ compared to the amplitude obtained when naively extrapolating the $k^{-11/3}$ behaviour down to the energy injection scale $K=1$, as shown in Fig. \ref{comparison}. This is an important point when we compare  the amplitude of the GW signal with previous estimates in the literature. Note that expression (\ref{velocityspectrum}) is analytic for $k\ra 0$, which is required for causally generated, incompressible turbulence in the cosmological context \cite{CDturb}. 

In the inertial range, the characteristic velocity on a given scale $2\pi/k$ 
is  approximately given by~\cite{Davidson}
\be
\label{velocity}
v_k^2\sim k E(k)
\simeq 6 \pi {\cal C}_v \frac{\Omega_T}{\Omega_{\rm rad}}K^{-2/3}
           ~~~~~~\mbox{ for}~K\gtrsim 3 \,.
\ee
Thus, the characteristic velocity on the energy injection scale, which is related to the total kinetic energy in the turbulence, is:
\be
v_L^2\simeq 6\pi \mathcal{C}_v \frac{\Omega_T}{\Omega_{\rm rad}}=4\pi \mathcal{C}_v \vev{v^2}\,.
\label{vL}
\ee

\subsection{The magnetic field power spectrum}
\label{sec:magnetic}

The power spectrum of the MHD processed magnetic field is closely related to the one of the turbulent velocity field. Here we treat the two sources exactly on the same footing. Since the energy density of a cosmological magnetic field scales like radiation,  we can use Eq.~(\ref{definition}) to evaluate the GW spectrum sourced by the magnetic field, provided that we define the normalized magnetic field vector
\be
b_i=\sqrt{\frac{3}{16\pi \rho_{\rm rad}}}B_i\,,
\label{bnormal}
\ee
so that the transverse traceless $(TT)$ part of the magnetic field energy 
momentum tensor is
 \be
\left[T^B_{ij}(\bx,t)\right]^{(TT)} =
\left[\frac{B_i(\bx,t)B_j(\bx,t)}{4\pi}\right]^{(TT)} =
\frac{4}{3}\rho_{\rm rad}(t)\left[b_i(\bx,t)b_j(\bx,t)\right]^{(TT)} =
 (\rho+p) \tilde\Pi^B_{ij}(\bx,t)\,.
\ee
The normalized magnetic field $b_i$ is equivalent to the dimensionless turbulent velocity field $v_i$. We define a parameter analogous to Eq.~(\ref{OmTparameter}), given by the ratio of the magnetic field energy density to the radiation energy density
\be
\vev{b^2}=\frac{3}{2}\frac{\Om_B}{\Om_{\rm rad}}=2\frac{\rho_B}{\rho+p}\,, \qquad \qquad {\rm with} \qquad \rho_B=\frac{\vev{B^2}}{8\pi}\,.
\ee
In the following we assume equipartition between the magnetic and turbulent energy densities at the time when the latent heat is released, therefore $\vev{v^2}\simeq \vev{b^2}$ (however, the scaling of the GW spectra with the source energy density is kept explicit). The slope of the high frequency tail of the magnetic power spectrum in fully developed MHD turbulence is not precisely  known: it could be of the Kolmogorov type, or it could satisfy the Iroshnikov-Kraichnan \cite{Iroshnikov-Kraichnan} or Goldreich-Sridhar \cite{Goldreich-Sridhar} spectral slopes. While in the presence of a strong background magnetic field, the spectrum of the field component perpendicular to the background field is of the Iroshnikov-Kraichnan type according to Refs. \cite{Maron,Muller,Gogobe}, in the isotropic case, the relevant one in cosmology, the simulations of Ref.~\cite{Muller} indicate a Kolmogorov-type slope. However, to diversify the treatment of the magnetic source from the turbulent one, we choose to consider the Iroshnikov-Kraichnan spectrum; the GW spectrum resulting from a Kolmogorov magnetic field is not very different and can be readily derived from the turbulent one. 

The low frequency tail of the spectrum is determined by causality and by the fact that $\bf B$ is divergence free. Like for turbulence, we use the interpolating formula from Ref.~\cite{Karman} to find the equal time magnetic power spectrum 
\bea
\vev{b_{i}(\bk,t)b_{j}^*(\bq,t)}&=&(2\pi)^3 \de(\bk-\bq)\,P_{ij}\,P_b(k,t)\,, \label{bequaltime}\\
P_b(k,t)&=&\frac{3}{2}\,\mathcal{C}_b \,\frac{\Om_B}{\Om_{\rm rad}} \,L^3\,\frac{K^2}{(1+K^2)^{11/4}}\times 
\left\{\begin{array}{ll}
1 & {\rm for}~0\leq K\leq \frac{L}{\lambda} \\
0 & {\rm for}~K\geq \frac{L}{\lambda} \, .
\label{bspectrum}
\end{array}\right.
\eea 
The normalization constant $\mathcal{C}_b=\frac{7}{16\pi^{3/2}}\frac{\Gamma(3/4)}{\Gamma(1/4)}\approx 0.0265$ is calculated in the same way as in Eq.~(\ref{kinen}).  The magnetic field and the turbulent flow being generated by the same physical process, namely bubble nucleation and collision, we assume that they share the same correlation scale $L$ and Kolmogorov microscale $\lambda$. 

\subsection{Freely decaying turbulence}
\label{sec:freedecay}

The turbulence stirring time is given by the duration of the phase transition which is typically much shorter than one Hubble time. Calling $\tin$ the time at which the phase transition starts,
one has $\beta^{-1}\ll \mathcal{H}_{\rm in}^{-1}$. As discussed in Section \ref{sec:turbuinitial}, the typical time interval over which turbulence is established is the eddy turnover time $\tau_L$.  Moreover, Kolmogorov  turbulence can be generated only if $\tau_L\leq \mathcal{H}_{\rm in}^{-1}$. This condition translates into a lower bound for the turbulent fluid velocity, in terms of the phase transition parameters: 
\be
v_f\gtrsim (\mathcal{H}_{\rm in}/\beta) v_b\,.
\ee
If $\tau_L\leq \mathcal{H}_{\rm in}^{-1}$, turbulence sets in after a short interval of time $\Det\equiv \tau_L \leq \tin$. 
After the completion of the phase transition (once 
bubbles have percolated), the stirring is over and the turbulence enters the free decay regime, \emph{i.e.} 
the total kinetic energy is dissipated (see for example \cite{Davidson}). The physical quantities 
appearing in the turbulent (\ref{velocityspectrum}) and magnetic (\ref{bspectrum}) spectra are comoving, but they also have an 
additional `absolute' time dependence due to the evolution of the turbulent 
cascade and the decay of the total energy. As already mentioned in the introduction, the existence of a maximal
correlation length in the universe (the horizon) and the consequent Batchelor spectrum (\emph{i.e.} $k^2$ at large scales), 
imply that $(\Omega_T/\Omega_{\rm rad})L^5$ is constant in time~\cite{Loi,LLt,CDturb}. 
According to this, we assume the following laws for the growth of the 
correlation scale and the decay of the kinetic energy:
\bea  \label{e:Leta}
 L(t)&=&L_*\left(\frac{t-\tin}{\tau_L}\right)^\gamma\,, \qquad \ga>0 \\
\label{e:OmTeta}
\frac{\Omega_T}{\Omega_{\rm rad}}(t)&=&\frac{\Omega_{T*}}{\Omega_{\rm rad*}}
  \left\{\begin{array}{ll}
\frac{t-\tin}{\tau_L}\,, & \tin\leq t\leq \tin+\tau_L\,,\\
 \left(\frac{\tau_L}{t-\tin}\right)^{5\gamma}\,, & t\geq \tin+\tau_L \,.
\end{array}\right. 
\eea
When the phase transition starts, at $\tin$, both the correlation length and the kinetic energy vanish. This insures that the source is continuous in time. At a time $t_*=\tin+\tau_L$, after the completion of the phase transition, turbulence is fully developed. The stirring scale is given by $L_* $, and 
$\omt$ is the total kinetic energy in the turbulent fluid, normalized to the radiation 
energy at time $t_*$. We assume that the energy cascade responsible for the Kolmogorov spectrum starts at this stage. At times $t\geq t_*=\tin+\tau_L$ turbulence enters the free decay phase, and the correlation scale and the kinetic energy evolve following the condition $(\Omega_T/\Omega_{\rm rad})L^5=$constant. 

The simple linear interpolation between these two behaviors given in (\ref{e:OmTeta}) has been introduced to mimic the turning on of the source in the bubble collision case, inferred from numerical simulations, see \cite{Huber:2008hg}. The linear increase of the magnetic energy density in MHD turbulence has also been observed in simulations \cite{Cho:2008ah}. After this initial phase, for Kolmogorov 
turbulence which we shall adopt here, the energy decay law
infers the value $\gamma=2/7$ (see \eg \cite{Davidson} and the introduction). For generality, we keep $\gamma$ unspecified  in the 
analytical formulae. When numerical results are presented, we substitute the 
value $\gamma=2/7$. 

In order to show explicitly its time evolution, we re-express  the velocity power spectrum 
Eq.~(\ref{velocityspectrum}) in terms of the time-independent variable 
$K_*=k L_*/2\pi$. Introducing the dimensionless time variable 
$$y=\frac{t-\tin}{\tau_L}\, ,$$ 
consequently 
$L(t)=L_*\,y^{\gamma}$, and using $K=K_*\,y^{\gamma}$ we obtain:
\bea
\label{vspectrumy}
{P_v\left(K_*,y\right)}=\frac{3}{2} \ \mathcal{C}_v 
\frac{\Omega_{T*}}{\Omega_{\rm rad*}} \, {L_*^{3}} \, 
    \frac{K_*^2}{(1+K_*^2y^{2\gamma})^{17/6}} \times
  \left\{\begin{array}{ll}
    y^{5\gamma+1} & \mbox{if}~0\leq y\leq 1~{\rm and}~0\leq K_*\leq \frac{L_*}{\lambda(y)} \nonumber \\
     1 & \mbox{if}~y\geq 1~{\rm and}~0\leq K_*\leq \frac{L_*}{\lambda(y)}  \nonumber \\
     0 & \mbox{if}~K_*\geq \frac{L_*}{\lambda(y)}
\end{array} \right.
\eea

\begin{figure}[htb!] 
\begin{center}
\includegraphics[width=8cm]{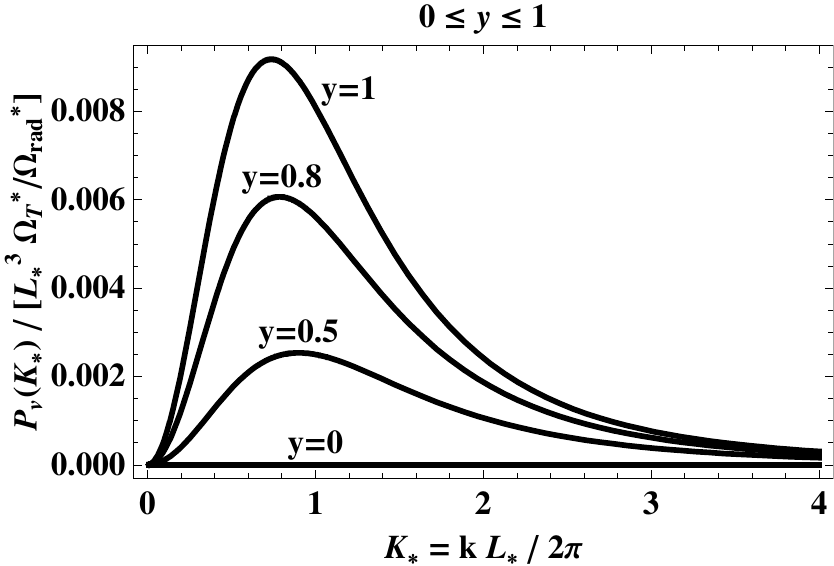}
\includegraphics[width=8cm]{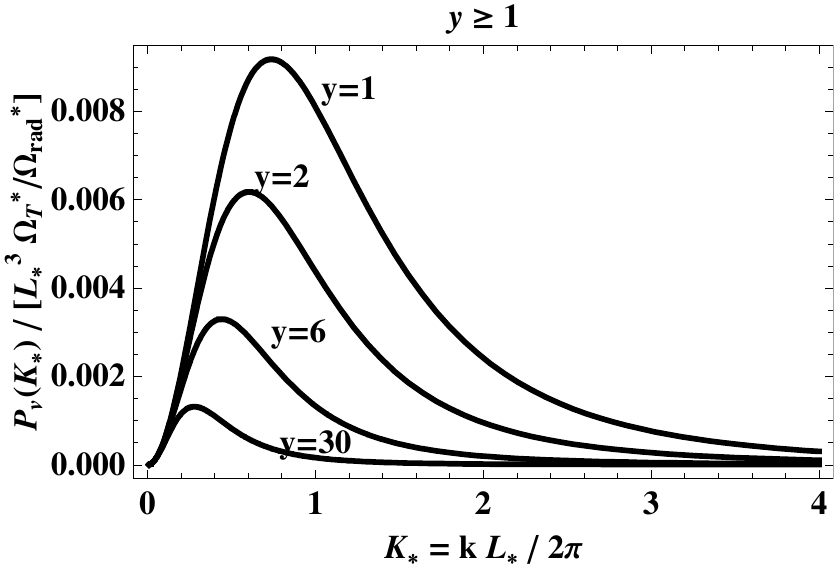}
\caption{\label{turbulent_velocity_spectrum}
\small The normalized velocity power spectrum as a
function of wavenumber $K_*$ for different times. Left: the phase in which 
the turbulence is developing, $0\leq y\leq 1$. Right: the phase of free 
decay, $y\geq 1$. The Kolmogorov microscale is outside the plot range (\cf end of section \ref{sec:howlong}).}
\end{center}
\end{figure}

\begin{figure}[htb!]
\begin{center}
\includegraphics[width=8.1cm]{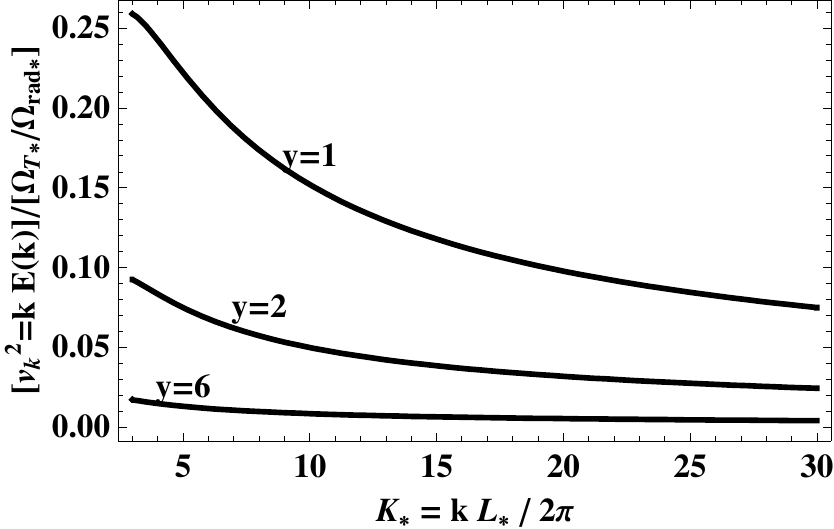}
\includegraphics[width=8.1cm]{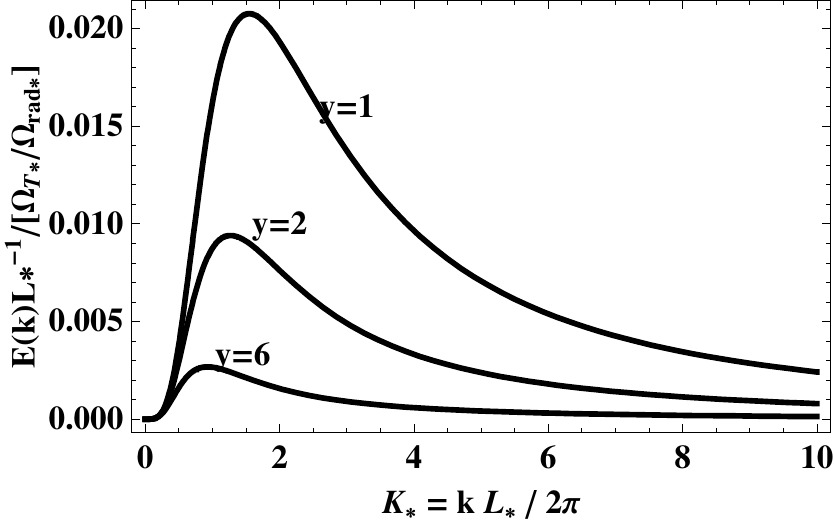}
\caption{\label{energy_spectrum} \small Left: the characteristic velocity 
$v_k^2\sim k E(k)=4\pi K_*^3P_v(K_*)/L_*^3$  as a function of wavenumber at 
different times in the inertial range $K_* \gtrsim 3$ during the free decay
phase, $y\geq 1$.
Right: the kinetic energy $E(k)/L_*=2 K_*^2P_v(K_*)/L_*^3$. The Kolmogorov microscale is outside the plot range. }
\end{center}
\end{figure}

The time-dependence of $P_v(K_*,y)$  is shown in Fig.~\ref{turbulent_velocity_spectrum}. The initial phase, $0\leq y\leq 1$ is inserted so that the source of GW increases smoothly from zero. As already discussed, the continuity of the source at initial time is an important issue for the resulting GW spectrum \cite{thomas}. However, since the duration of this initial phase is short, we assume that the details of the source spectrum during this phase are not relevant and we do not model them in any detail. In particular, we do not expect the inertial range $K_*^{-11/3}$ for $K_*\gtrsim 3$ to be already developed in this initial phase, because the real energy cascade has not yet started. Nevertheless, for simplicity we keep the same form of the spectrum as a function of $K_*$ in the two phases: the turbulent free decay phase, $y\geq 1$, is actually the most relevant one for the GW generation. During this phase, the large scale part of the power spectrum $K_*\ll 1$ remains constant. We also show the characteristic velocity and the kinetic energy as functions of $K_*$ for different times in Fig.~\ref{energy_spectrum}.

In the following we assume that the turbulent magnetic field also undergoes the same time decay as the turbulent velocity field. Even though the precise decay law of the magnetic field energy density is not known in general for MHD turbulence, both analytical \cite{olesen,shiro,son,Campanelli:2007tc} and numerical \cite{biskamp,Christensson:2000sp,Banerjee:2004df} analyses seem to agree with the fact that the magnetic field power spectrum is persistent on large scales. Together with the condition that the spectrum, in a cosmological setting, should be of the Batchelor type (\ie $k^2$ at large scales) due to the presence of a cosmological horizon, constancy in time at large scales entails the existence of a conserved quantity analogous to Loitsyansky's invariant: $(\Om_B/\Om_{\rm rad})L^5=$constant. This gives the same decay as for the turbulent flow. Equivalent scaling laws (once generalized to the Batchelor case) are obtained from the argument of self-similarity \cite{olesen,Christensson:2000sp} and direct cascade \cite{son,Banerjee:2004df}. Therefore, we also assume in the magnetic case, 
\bea
L(t)&=&L_*y^\gamma \nonumber \\
\nonumber
\frac{\Omega_B}{\Omega_{\rm rad}}(t)&=&\frac{\Omega_{B*}}{\Omega_{\rm rad*}}
  \left\{\begin{array}{ll}
y& 0 \le y \le 1\\
 y^{-5\gamma} & y\geq 1\,,
\end{array}\right.  \\
{P_b\left(K_*,y\right)}&=&\frac{3}{2} \ \mathcal{C}_b 
\frac{\Omega_{B*}}{\Omega_{\rm rad*}} \, {L_*^{3}} \, 
    \frac{K_*^2}{(1+K_*^2y^{2\gamma})^{11/4}} \times
\left\{\begin{array}{ll}
    y^{5\gamma+1} & \mbox{if}~0\leq y\leq 1~{\rm and}~0\leq K_*\leq \frac{L_*}{\lambda(y)} \nonumber \\
     1 & \mbox{if}~y\geq 1~{\rm and}~0\leq K_*\leq \frac{L_*}{\lambda(y)}  \nonumber \\
     0 & \mbox{if}~K_*\geq \frac{L_*}{\lambda(y)} \,.
\end{array} \right.
\eea
Our argument for assuming this particular form of the time decay in MHD turbulence relies on approximative, analytical considerations, and given the high non-linearity of the problem it is conceivable that only numerical simulations will be able to find the correct scaling. For simplicity, and in order to be able to proceed with our analytical estimate, we are forced to make the above mentioned, rather crude assumptions for the scaling. Nonetheless, we would like to stress here once again the importance of the presence of a causal horizon, unavoidable in the cosmological setting. This prevents the formation of long range correlations at least beyond the horizon scale, a feature which certainly affects the decay law and that can not be accounted for in numerical MHD simulations which
go on for times larger than the box size. 

\subsection{How long does turbulence last?}
\label{sec:howlong}

In this section we confirm that turbulence is generated during a phase 
transition, and we determine when it ends according to the free decay 
picture described above. For this we  evaluate the Reynolds number defined in
Eq.~(\ref{Reynold_number}) at the energy injection scale $L$ corresponding to 
$K=kL/2 \pi =1$. We distinguish the physical length with a subscript $_p$ 
\be 
L_{p}(T)= L(y)  \frac{T_0}{T} \left(\frac{g_0}{g(T)}\right)^{1/3}
\ee
from the comoving scale $L(y)=L_* y^{\gamma}$.
The kinematic viscosity $\nu(T)$ is derived in 
Appendix~\ref{Appen:viscosity}:
\begin{eqnarray}
\nu(T)\approx   \left\{\begin{array}{ll}
 {22} \ {T^{-1}} &T \gtrsim 100\mbox{ GeV}\\
5  \ 10^8 \ {\mbox{GeV}^4} \ {T^{-5}} &T \lesssim 100 \mbox{ GeV }\\
2  \ 10^9 \ {\mbox{GeV}^4} \ {T^{-5}} &T \lesssim 100 \mbox{ MeV }\\
\end{array}\right.
\label{nudefinition}
\end{eqnarray}
The jumps in the viscosity introduce an uncertainty in the evaluation of the 
parameters, for which we can only give the correct order of magnitude. 

\begin{figure}[htb!]
\begin{center}
\includegraphics[width=10cm]{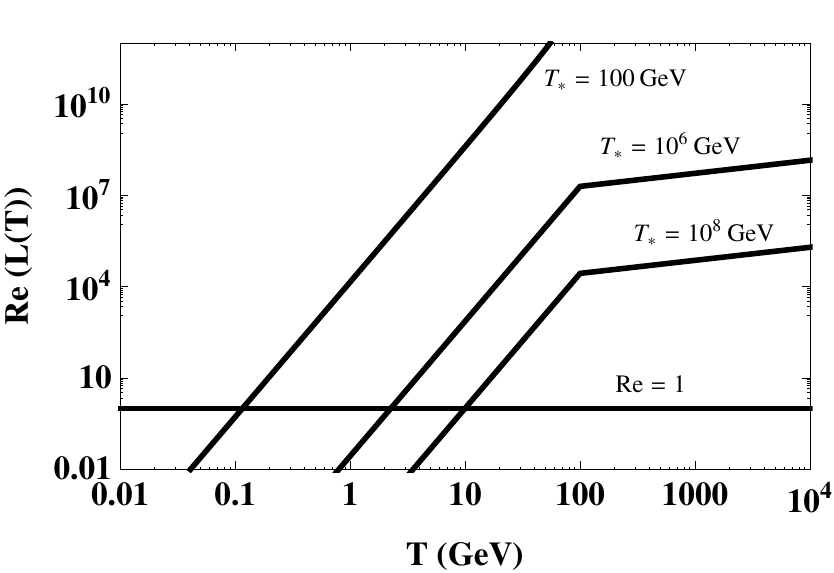}
\caption{\label{Reynold} \small Evolution of the Reynolds number  at the energy 
injection scale $L_p(T)$ as a function of temperature,  assuming a phase 
transition with $\beta/\HH_*=100$, $\omt=2/9$ and $\gamma=2/7$, for three 
phase transition temperatures: $T_*=100, 10^6$ and $10^8$ GeV. The temperature at 
which the Reynolds number decays below 1 (horizontal line) represents 
approximatively the end of 
the turbulence: this typically happens at temperatures $\sim 100$ MeV, 2 GeV 
and 8 GeV respectively.}
\end{center}
\end{figure}

For $v_L\equiv v_k(k=2\pi/L)$, we use the relation (\ref{velocity}). 
Eq.~(\ref{velocity}) is valid only during the cascade, the free decay phase 
$y\geq 1$, and in the inertial range $K\gtrsim 3$. To estimate the Reynolds 
number, we extrapolate it to $K=1$, thus making a small error. However, the 
following estimate is valid only for times $y\geq 1$. We have
\be
v^2_L(y)\sim 6\pi{\cal C}_v\frac{\Omega_{T*}}{\Omega_{\rm rad*}} y^{-5\gamma}\,,
   ~~~\mbox{for}~~y\geq 1\,.
\label{vLsquared}
\ee
Inserting this in (\ref{Reynold_number}) we obtain for temperatures 
$T\leq T_*$ corresponding to $y\geq 1$
\bea
\Re(L(T)) &=& \Re(L_*) \frac{T_*}{T} \, \frac{\nu(T_*)}{\nu(T)} \, 
y^{-\frac{3}{2}\gamma} \left(\frac{g_*}{g(T)}\right)^{1/3} \quad \mbox{with} 
\label{reynoldsT}\\
\Re(L_*)&=&\sqrt{6\pi{\cal C}\frac{\Omega_{T*}}{\Omega_{\rm rad*}}} 
\frac{L_*}{\nu(T_*)} \frac{T_0}{T_*}\left(\frac{g_0}{g_*}\right)^{1/3} \,.
\label{reynoldsT*}
\eea
$\Re(L(T))$ is plotted for different values of $T_*$ in  Fig.~\ref{Reynold}. 
If we consider the EW phase transition, for which $T_*\sim 100\gev$, 
and we fix $\beta/\HH_*=100$, $\Om_{T*}/\Om_{\rm rad*}=2/9$ corresponding to 
$v_f=c_s$, $v_b=0.87$ corresponding to detonations (see Sec.~\ref{sec:turbuinitial}), and $\gamma=2/7$, we find the 
value
\be
\Re(L(T_*=100\gev)) \sim 10^{13}\,.
\label{reynolds100}
\ee
This confirms that turbulence develops once the primordial fluid is stirred 
on the scale $L_*=2 v_b/\beta$.
In Appendix~\ref{Appen:viscosity} we also derive
\be
{\rm P_m} =\frac{\nu}{\mu}\simeq 10^{12}\left(\frac{\rm GeV}{T}\right)^4 \,, \qquad
1~{\rm MeV}< T< 100~{\rm GeV}\,,
\ee
so that
\bea
{\rm R_m}(L(T)) = \Re(L(T)){\rm P_m}(T)  &\simeq& 10^{12} \,\Re(L(T))
\left(\frac{\rm GeV}{T}\right)^4 \,, \\
 {\rm R_m}(L(T_*=100\gev)) &\sim& 10^{17}\,.
\eea

\begin{figure}[htb!]
\begin{center}
\includegraphics[width=10cm]{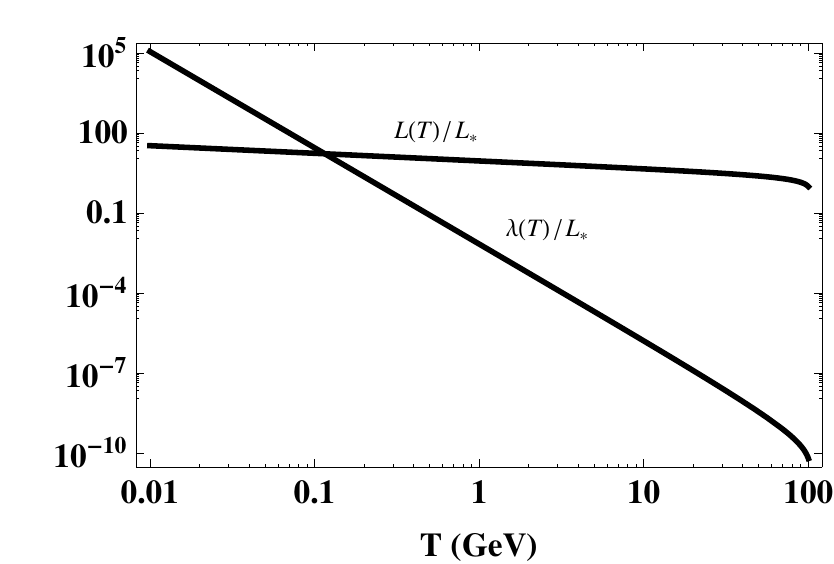}
\caption{\label{scales} \small Evolution of the stirring scale 
$L(T)$ and the Kolmogorov microscale $\lambda(T)=L(T)/[\Re(L(T))]^{3/4}$ with 
temperature, for $T_*=100$ GeV, $\beta/\HH_*=100$, $\omt=2/9$ and $\gamma=2/7$.}
\end{center}
\end{figure}

We now want to determine the wavelength up to which the cascade is present, and the time up to which MHD turbulence persists. The 
turbulent cascade stops when viscosity becomes important, and this happens at 
scales smaller than the Kolmogorov microscale $\la$, defined by
\be
\Re(\la)=\frac{v_{\la} \,\la_p}{\nu}\equiv 1 ~~~~\mbox{ hence }~~~~
\frac{L}{\la}=\frac{v_{\la}}{v_{L}}\,\Re(L)=\Re(L)^{3/4}\,,   \label{Loverla}
\ee
where for the last equality we use (\ref{velocity}). Therefore, the Kolmogorov 
microscale also grows during free decay, according to: 
\bea
\lambda(T)=\lambda_{*} \left(\frac{T}{T_*} \, \frac{\nu(T)}{\nu(T_*)}\left(\frac{g(T)}{g_*}\right)^{1/3}\right)^{3/4}\,y^{\frac{17}{8}\gamma}\,.
\label{laT}
\eea
The temperature dependence of $\lambda(T)$ and $L(T)$ is illustrated in 
Fig. \ref{scales}. The ratio of the initial scales determining the extension 
of the turbulent inertial range is $L_{*}/\la_*=\mathcal{O}(10^{10})$, for $T_*=100\gev$, $\beta/\HH_*=100$, 
$\Om_{T*}/\Om_{\rm rad*}=2/9$, $v_b=0.87$ and $\gamma=2/7$.

Since $\la(T)$ grows faster than $L(T)$, there always exists a temperature at 
which the two scale cross, as shown in Fig. \ref{scales}. We define the end 
of turbulence when the entire inertial range ($K\gtrsim 3$) is dissipated, 
namely when the dissipation scale has grown to reach
\be
\frac{L(T_{\rm fin})}{\lambda(T_{\rm fin})}=3 ~~~~\Rightarrow~~~~
  \Re(L(T_{\rm fin}))=3^{4/3}\,.
\label{Tfindefinition}
\ee
If turbulence starts at $T_*=100\gev$, using again the values $\beta/\HH_*=100$, 
$\Om_{T*}/\Om_{\rm rad*}=2/9$ and $\gamma=2/7$, we find 
$T_{\rm fin}\simeq 120~\mev$. Therefore, turbulence acts as source of 
GWs for many Hubble times. 

Once turbulent motions are 
dissipated, the magnetic field fluctuations at scales larger than the 
dissipation scale remain frozen in the primordial plasma. Hence, in principle, 
the magnetic field continues acting as a source of GW also after the end of 
turbulence. However, we neglect this extra contribution to the GW spectra 
since it will not affect the spectrum at the interesting scales, around the 
peak, where the signal may be visible. These scales in fact have entered the 
horizon well before the end of turbulence and further GW production is 
strongly suppressed at later times.  

\subsection{Time de-correlation of the spectrum of magneto-hydrodynamical turbulence}
\label{sec:decorrelation}

In order to evaluate the GW power spectrum generated from MHD turbulence, the 
equal time velocity and magnetic field power spectra given in Eqs.~(\ref{velequaltime}, \ref{velocityspectrum}) and Eqs.~(\ref{bequaltime}, \ref{bspectrum}) are not enough. We need to know the unequal time 
velocity power spectrum, 
\be
\vev{v_{i}(\bk,t_1)v_{j}^*(\bq,t_2)}=(2\pi)^3\de(\bk-\bq)\,P_{ij}\,P_v(k,t_1,t_2)\,,
\label{velunequaltime}
\ee 
and equivalently for the magnetic field. To model the unequal time 
velocity power spectrum, we multiply the equal time one (\ref{velocityspectrum})
with the exponential time de-correlation proposed by Kraichnan in~\cite{Kraichnan}  
and also used in 
\cite{Tinaturb}. The temporal de-correlation of the velocity field  described in~\cite{Kraichnan} 
operates only in the inertial range and during the cascade, which in our case 
means $K\gtrsim 3$ and $t\geq t_*$. The characteristic de-correlation time scale 
is given by the eddy turnover time: $\tau_\ell=\ell/(2v_\ell)$, for a 
characteristic eddy of size $\ell=2\pi/k$ well in the inertial range. To model 
the de-correlation Kraichnan proposes a Gaussian functional form,
\be
g(t_1,t_2)=\exp(-\pi(t_1-t_2)^2/(4\tau^2_\ell))\,.
\label{Krai}
\ee
In our approach, the magnetic field undergoes the same de-correlation 
as the turbulent field. 

In the case of freely decaying MHD turbulence under consideration here, the eddy turnover time $\tau_\ell$ is itself time-dependent, through the characteristic velocity on the same scale, $v_\ell$. Using the formula of the characteristic velocity in the inertial range (\ref{velocity}), and the evolution equations for the kinetic energy and the correlation scale (\ref{e:OmTeta}) and (\ref{e:Leta}) we obtain
\be
\tau_\ell(y)=\frac{\ell}{2v_\ell}=\tau_\ell^* \,\,y^{\frac{17}{6}\gamma}~~~~~\mbox{where}~~~~~
\tau_\ell^*=\frac{L_*^{\frac{1}{3}}\,\, \ell^{\frac{2}{3}}}{2\sqrt{6\pi\mathcal{C}_v\omt}} \,,
\ee 
and the above equation is valid only during the cascade $y\geq 1$ and $K\gtrsim 3$. 
Setting $y=(t_1-\tin)/\tau_L$, and $z=(t_2-\tin)/\tau_L$, we arrive at the 
following expression for the velocity unequal time power spectrum (the 
magnetic field spectrum is readily derived from the expression below, 
substituting $\mathcal{C}_v$ by $\mathcal{C}_b$, $\Om_T$ by $\Om_B$ and 
$17/6$ by $11/4$):
\bea
\label{Pvmixedtime}
P_v(K,y,z)&=&\frac{3}{2}\,\mathcal{C}_v\,\frac{\Omega_T}{\Omega_{\rm rad}}(y,z)\,
    L^3(y,z)\,\frac{K^2(y,z)}{[1+K^2(y,z)]^{17/6}}
 \label{Pv}\\
&\times&\left\{\begin{array}{ll}
1 & {\rm for}~K(y,z)<3 \\
\exp\left(-\frac{\pi}{4}(y-z)^2\big(\frac{\tau_L}{\tau_\ell(y,z)}\big)^2\right) &{\rm for} ~
   3<K(y,z)<\frac{L}{\lambda} ~{\rm and}~y,z\geq 1 \nonumber\\
0 & {\rm for}~K(y,z)>\frac{L}{\lambda}
\end{array}\right.
\eea
Here the notation $\frac{\Omega_T}{\Omega_{\rm rad}}(y,z)$ (and so on) is to remind that the variables $K$, $L$, $\Om_T/\Om_{\rm rad}$ and $\tau_\ell$ depend on time. In principle, we need to specify at which time these variables have to be evaluated. This choice must satisfy the constraint that the unequal time power spectrum (\ref{velunequaltime}) is symmetric under the exchange of $t_1$ and $t_2$. As explained in the next section, we avoid the problem by modeling de-correlation of the Kraichnan type directly in the anisotropic stresses of the source.

\section{The anisotropic stress power spectrum}
\label{sec:anisotropic}

In order to determine the GW energy density power spectrum generated by 
the MHD turbulent source, we need to calculate the anisotropic 
stress power spectrum (\ref{Pispectrum}) at different times, substitute it into 
Eq.~(\ref{definition}), and evaluate the time integral. 

The tensor anisotropic stress is the transverse traceless part of the energy 
momentum tensor 
$\tilde \Pi_{ij}(\bk,t)=(P_{il}P_{jm}-1/2P_{ij}P_{lm})\tilde T_{lm}(\bk,t)$, where we denote the 
dimensionless energy momentum tensor with a tilde, see Eq.~(\ref{TTT}). The part of the energy momentum tensor of 
our sources which contributes to the anisotropic stress is given by $\tilde T^{(T)}_{lm}(\bx,t)=v_l(\bx,t) v_m(\bx,t)$ 
and $\tilde T^{(B)}_{lm}(\bx,t)=b_l(\bx,t) b_m(\bx,t)$. The anisotropic stress power spectrum is then 
\be
\vev{\Pi_{ij}(\bk,t_1)\Pi_{ij}^*(\bq,t_2)}=(2\pi)^3\de(\bk-\bq)\Pi(k,t_1,t_2)=
\mathcal{P}_{abcd}\vev{T_{ab}(\bk,t_1)T_{cd}^*(\bq,t_2)}\,,
\label{anisostress}
\ee
 where we omit the tildes for simplicity and
 \be
 \mathcal{P}_{abcd}=\left(P_{ia}P_{jb}-\frac{1}{2}P_{ij}P_{ab}\right)({\bf{k}})
 \left(P_{ic}P_{jd}-\frac{1}{2}P_{ij}P_{cd}\right)({\bf{q}})\,.
 \ee
  For the turbulent source, one has 
 \bea
&& \vev{\Pi_{ij}(\bk_1,t_1)\Pi_{ij}^*(\bk_2,t_2)} = \nonumber \\
&&\hspace*{-0.3cm} \mathcal{P}_{abcd}\int 
\frac{d^3p}{(2\pi)^3} \int \frac{d^3q}{(2\pi)^3}
  \vev{v_a(\bk_1-\bp,t_1) v_b(\bp,t_1) v_c^*(\bk_2-\bq,t_2) v_d^*(\bq,t_2)}\,.
\label{Pifour}
 \eea
In order to proceed analytically, we assume that we can decompose the four 
point correlation function into products of the power spectra, using Wick's 
theorem like for a Gaussian random field. Since turbulence is not truly Gaussian,
this is of course not strictly correct but it is usually adopted as a 
reasonable approximation to close the hierarchy (\emph{i.e.} to avoid using equations involving higher order correlators).
We refer to \cite{GW1mag} and Appendix~\ref{app:turb} for details of the 
determination of $\Pi(k,t_1,t_2)$. The bottom line is that the anisotropic 
stress power spectrum is given by the convolution of the unequal time velocity 
power spectrum, \cf Eq.~(\ref{a:Pispectrumint}). This quantity is in principle determined once we know how the 
source behaves in time. In our case it is given by the Kraichnan de-correlation 
model, once we have chosen an appropriate way of symmetrizing in time, see 
Eq.~(\ref{Pvmixedtime}).   

On the other hand, the anisotropic stress power spectrum should, by definition,
be a positive kernel, \ie such that 
\be
\int dt_1 \int dt_2\, \Pi(k,t_1,t_2) f(t_1)f^*(t_2) \geq 0 \qquad \forall\,\, f(t)\,.
\label{poskernel}
\ee
This follows simply from the definition of the power spectrum, 
Eq.~(\ref{anisostress}), making use of the inequality 
\be
\vev{\left| \int dt_1\, \Pi_{ij}(\bk,t_1) f(t_1) \right|^2} \geq 0\,.
\ee
Comparing Eq.~(\ref{poskernel}) with the definition of the GW spectrum 
Eq.~(\ref{definition}), we see that in the GW case the function $f(t)$ 
is replaced by the Green function of the GW wave equation, 
$f(t_1)=\cos(kt_1)/t_1$ or $\sin(kt_1)/t_1$ (the factor $1/t_1$  is absent for 
the short lasting case). Since the trigonometric functions are a complete 
basis, the property (\ref{poskernel}) is automatically satisfied, if it holds
for the Green function with arbitrary values of $k$. Therefore, inserting the 
unequal time source power spectrum Eq.~(\ref{Pvmixedtime}), which accounts for 
Kraichnan de-correlation, in the convolution given by Eqs.~(\ref{Pifour}) and~(\ref{a:Pispectrumint}), and performing 
the integration over the momenta, should give back an anisotropic stress 
power spectrum which is a positive kernel. However, we have tried different 
ways of symmetrizing Eq.~(\ref{Pvmixedtime}) without succeeding in obtaining 
a positive kernel. This may be related to the inaccuracy of our analytical 
estimates, to our choice of symmetrization, or it could be an indication that 
either Wick's theorem or free decay together with Kraichnan de-correlation, although reasonable assumptions for the evolution of MHD turbulence,  
are not entirely appropriate for the source we are considering. Only a full numerical 
simulation of the turbulent and magnetic fields could help to understand the
shortcomings of our approach.

The same problem can arise when trying to obtain the source power 
spectrum by Fourier transforming the space correlation function, but this is 
circumvented since the source is modeled directly in Fourier space (the velocity power 
spectrum is well known in the MHD turbulent case). However, in order to 
calculate the GW spectrum one has to evaluate also the time Fourier transform: this 
is not a common procedure and no general solution is given in the literature. The 
problem is worsened by the fact that we want to account for the free decay of 
turbulence. If the only time dependence of the turbulent power spectrum was
the Kraichnan de-correlation, which is Gaussian in the time difference, 
the spectrum resulting after time Fourier transforming would be positive. 
However, the free decay introduces an additional absolute time dependence which
is more difficult to handle. Moreover, having to evaluate the anisotropic 
stress power spectrum further increases the difficulty of finding the correct 
time behaviour that would provide a positive kernel. Note that the same 
problem arose in the analytical evaluation of the GW signal coming from 
bubble collisions \cite{Caprini:2007xq}. 

In order to proceed, we model the source in such a way, that 
$\Pi(k,t_1,t_2)$ is a positive kernel by construction. This is most easily 
done directly for the anisotropic stress power spectrum. In previous 
analyses~\cite{Caprini:2007xq} we have already tackled this problem and 
proposed three forms for the unequal time anisotropic stress power spectrum 
which are positive kernels. In Sec.~\ref{sec:constant} we presented the 
incoherent (\ref{inco1}) and coherent (\ref{co1}) cases, for which the source 
is never correlated, respectively always correlated in time. We apply them in 
the following to the MHD turbulent source. We believe, however, that the top 
hat correlation introduced in Ref.~\cite{Caprini:2007xq} is the relevant 
one for MHD turbulence, since it best mimics the Kraichnan de-correlation. In
the top hat model one assumes that $\Pi(k,t_1,t_2)$ is correlated if $|t_1-t_2| <x_c/k$ 
and uncorrelated otherwise. Here $x_c$ is a parameter of order unity. We shall 
choose $x_c=1$ for our numerical results. Since $\Pi(k,t_1,t_2)$ has to be symmetric in 
$t_1$ and $t_2$ we set
\bea
\Pi(k,t_1,t_2)&=& \frac{1}{2} \left[ \Pi(k,t_1,t_1) \,\Theta(t_2-t_1)\,
\Theta\left(\frac{x_c}{k}-(t_2-t_1)\right)\right. \nonumber \\  &+& \left. 
\Pi(k,t_2,t_2) \,\Theta(t_1-t_2)\,\Theta\left(\frac{x_c}{k}-(t_1-t_2)\right) 
   \right]\,.
\label{tophatpi}
\eea
This way of correlating the source at unequal times is 
intermediate between the coherent and incoherent approximations; instead of being 
correlated at all times or only for $t_1=t_2$, here we account for the fact 
that longer wavelengths de-correlate at larger time differences. 
While the term $\Theta(t_1-t_2)$ is there
only to make the function symmetric and does not influence the time
continuity of the source, the term $\Theta(x_c/k-|t_1-t_2|)$ should in
principle be replaced by an exponential decay to keep the source continuous.
We have tried this, and a part from a much slower convergence of the numerical
integrals since the integrand oscillates rapidly for large values of 
$k|t_1-t_2|$, we found no difference in the final result. Moreover, inserting
(\ref{tophatpi}) in the integral (\ref{longlastingcase}), we find that the GW 
energy power spectrum is not given by the Fourier transform of the source and 
continuity does not affect the final spectrum in this case (\cf \cite{thomas}).

Assuming that the Kraichnan time de-correlation trivially extends from the 
turbulent velocity field to the anisotropic stress, it is clear that it 
gives a behaviour quite similar to the top hat de-correlation: according to
(\ref{Krai}), the source is no longer correlated for time intervals 
$|t_1-t_2|\gtrsim (2/\sqrt{\pi}) \tau_\ell$. Accounting for the fact that the 
eddy turnover time is simply the inverse of the characteristic frequency of 
the source $\omega_\ell=1/\tau_\ell$, and that the GW Green function selects the 
diagonal of the time Fourier transform of the source, for which $|\omega|=k$, 
we find that the Kraichnan de-correlation gives back the same condition as the 
top hat de-correlation, namely correlation is lost for time differences
\be
|t_1-t_2| \gtrsim \frac{2}{\sqrt{\pi}}\frac{1}{\omega_\ell}\simeq 
\frac{x_c}{k}\qquad {\rm with}~x_c\simeq 1\,. 
\ee
Even though the top hat case best reproduces turbulent de-correlation, in the 
following we evaluate the GW spectra also for the coherent and incoherent 
cases and compare the results. In order to proceed with the calculation, we 
now evaluate the equal time anisotropic stress power spectrum, 
see Eq.~(\ref{tophatpi}).

\subsection{The equal time anisotropic stress power spectrum for magneto-hydrodynamical turbulence}

\begin{figure}[htb!]
\begin{center}
\includegraphics[width=8cm]{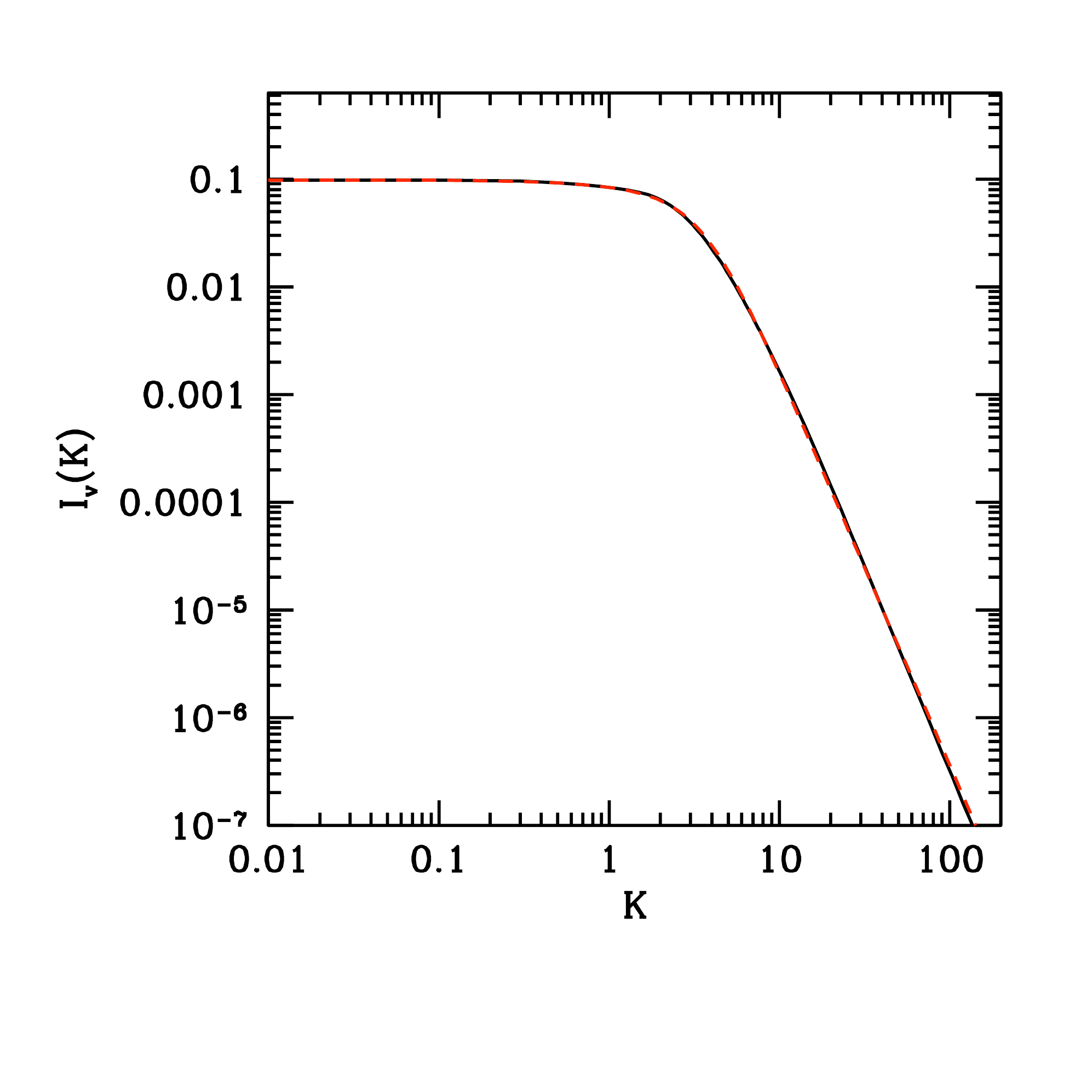}
\includegraphics[width=8cm]{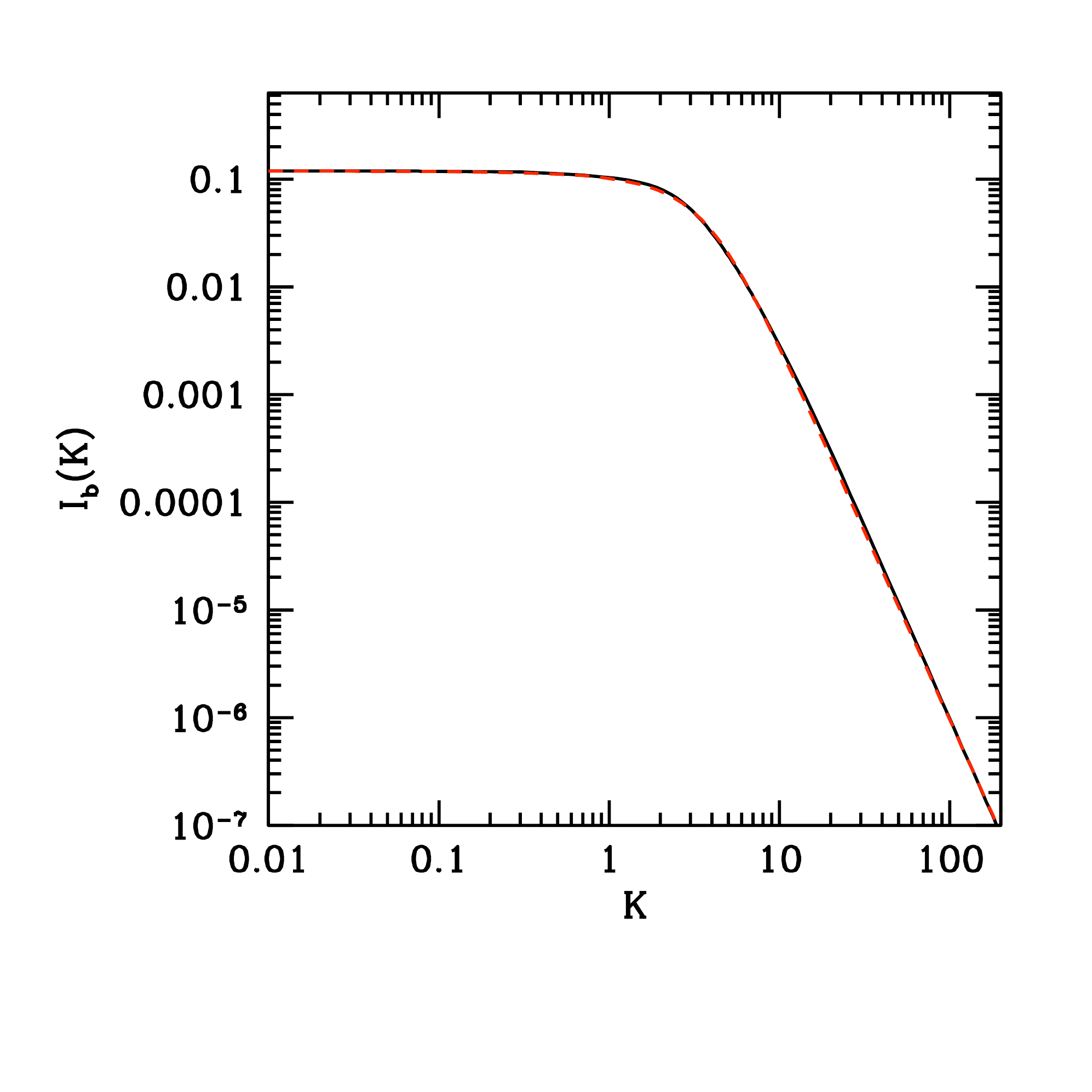}
\caption{\label{figfit} \small Black, solid lines: the integral in 
Eq.~(\ref{Icaltrue}), left: for the turbulence anisotropic stress, and right: 
for the magnetic field anisotropic stress, as a function of $K=K_*y^\gamma,$ 
together with their fits given in Eqs.~(\ref{Icalturb}) and (\ref{Icalmag})
(shown as red, dashed lines).}
\end{center}
\end{figure}

The equal time anisotropic stress power spectrum is given by the convolution 
of the equal time velocity and magnetic field power spectra multiplied by an angular dependence 
coming from the projector in Eq.~(\ref{Pifour}), see Eq.~(\ref{a:Pispectrumint}). As already mentioned above, 
we refer to \cite{GW1mag} and Appendix~\ref{app:turb} for a derivation. In 
terms of the variable $K=K_*y^\gamma$, using the velocity power spectrum given
in Eq.~(\ref{vspectrumy}), we find
\be
\Pi_v(K,y,y)=\frac{9}{2}\,\pi \, \mathcal{C}_v^2 \left(\frac{\Omega_T}{\Omega_{\rm rad}}(y)\right)^2L^3(y) \,\mathcal{I}_v(K,y,y) \,,
\ee
where $y=(t_1-\tin)/\tau_L$ and $\mathcal{I}_v$ is given by
\be
\mathcal{I}_v(K,y,y)=\int_0^\infty dQ\,\frac{Q^4}{(1+Q^2)^{17/6}}
\int_{-1}^1 d\chi\,(1+\chi^2)\frac{2K^2+Q^2(1+\chi^2)-4KQ\chi}{(1+ K^2-2\chi KQ +
   Q^2)^{17/6}}\,,
\label{Icaltrue}
\ee
with $Q=Lq/2\pi$, $\chi=\hat{k}\cdot\hat{q}$. The magnetic field anisotropic 
stress power spectrum is equivalent to the above expressions, substituting 
$\Om_T$ with $\Om_B$, $\mathcal{C}_v$ with $\mathcal{C}_b$ and the power 
law $17/6$ with $11/4$. The integrals cannot be performed analytically, so we 
solve them numerically and derive fits in terms of the variable
$K=K_*y^{\gamma}$. Within this approach we do not account for the small scale 
cutoff $L/\la$ when fitting the convolution. We explain below how this cutoff is taken into account. We find, for 
the turbulence and the magnetic field  respectively: 
\bea
\mathcal{I}_v(K_*,y,y)& \simeq &0.098\,\left[1+\left(\frac{K_* y^\gamma}{4}
      \right)^{4/3}+ \left(\frac{K_* y^\gamma}{3.3}\right)^{11/3}\right]^{-1} 
\label{Icalturb}\\
\mathcal{I}_b(K_*,y,y)& \simeq &0.12\,\left[1+\left(\frac{K_* y^\gamma}{4}
      \right)^{4/3}+ \left(\frac{K_* y^\gamma}{3.5}\right)^{7/2}\right]^{-1} 
\label{Icalmag}
\eea
The fits are shown in Fig.~\ref{figfit}.

\subsection{The final time of the magneto-hydrodynamical turbulent source} 
\label{sec:etafin}

As demonstrated in Sec.~\ref{sec:howlong}, MHD 
turbulence can act as a source of GWs for many Hubble times, while it 
undergoes free decay. The  source generally switches off at the end of the 
turbulence, when $\Re(L_p(T_{\rm fin}))=3^{4/3}$ (see Eq.~(\ref{Tfindefinition})).
 If $T_*=100$ GeV, we have found $T_{\rm fin}\simeq 120$ MeV. However, we have 
seen in Sec.~\ref{sec:howlong} that the Kolmogorov microscale grows in time, 
and for wavenumbers above this upper cutoff the source has decayed. 
This means that the final time of integration in Eq.~(\ref{longlastingcase}) 
must be defined as the minimum of $t(T_{\rm fin})$ and the time at which a given 
mode $k$ is equal to the upper cutoff. For later times, that mode $k$ is not 
generating GWs any longer; however, if the time at which $k$ is equal to the 
upper cutoff comes after the end of the turbulence, we should take the end of 
the turbulence $t(T_{\rm fin})$ as the final time of action of the source on 
the scale $k$. 
This is the way in which we include the upper cutoff $L/\lambda$ appearing 
in Eq.~(\ref{Pv}) in the calculation of the GW spectrum, even though we have 
neglected it for simplicity in the evaluation of the anisotropic stress power 
spectra (\ref{Icalturb}),~(\ref{Icalmag}). 

We introduce $t_k$ as the time at which $k=4\pi/\lambda(t_k)$, where the extra 
factor of 2 comes from the fact that  $\mathcal{I}_v(K_*,y,z)$, being the 
convolution of the velocity power spectrum (\ref{Pv}) (and equivalently for 
the magnetic field power spectrum), it will go to zero at twice the velocity 
power spectrum cutoff. Using the time evolution of the dissipation scale 
$\lambda$,  Eq.~(\ref{laT}), we find
\bea
\frac{t_k}{\tau_L}&=&\left[\frac{L_*}{\la_*}\frac{2}{K_*}
\left(\frac{t_*}{\tau_L}\right)^3\right]^{\frac{1}{17\gamma/8+3}}
~~~~{\rm if}~~T_*\leq 100\gev\,,~~T_{\rm fin}\geq 100~\mev
\label{etafin1} \\
\frac{t_k}{\tau_L}&=&\left[\frac{L_*}{\la_*}
   \frac{2}{K_*}\right]^{\frac{8}{17\gamma}}
~~~~{\rm if}~~T_{\rm fin}\geq 100 \gev\,,
\eea
where the different behaviour depending on the initial and final temperature 
is due to the different evolution of the viscosity, given in 
Eq.~(\ref{nudefinition}). Finally, the time at which turbulence at a given 
scale $k$ ends is $\tfin(k)$ given by
\be
\frac{\tfin(k)}{\tau_L}={\rm min}\left\{ \frac{t(T_{\rm fin})}{\tau_L}~,~\frac{t_k}{\tau_L}  \right\}
\label{etafin2}
\ee
where $t(T_{\rm fin}) \simeq T_0\, / \, (T_{\rm fin}\, H_0\sqrt{\Om_{\rm rad}})$ and  
$T_{\rm fin}$ denotes the end of the MHD turbulence by the dissipation of 
the entire Kolmogorov range, defined in Section~\ref{sec:howlong}. For 
the usual set of values $T_*=100~\gev$, $\beta/\HH_*=100$, 
$\Om_{T*}/\Om_{\rm rad*}=2/9$, $v_b=0.87$ and $\gamma=2/7$, we find 
\bea 
\frac{\tfin(k)}{\tau_L}\simeq\left\{\begin{array}{ll}
2\times10^{4} & \mbox{for}~K_*\lesssim 0.07 \\
\left(\frac{2\,\,10^{14}}{K_*}\right)^{\frac{28}{101}} & \mbox{for}~K_*\gtrsim 
0.07 \,.
\end{array} \right.
\eea

\section{The gravitational wave spectrum \label{s:GWspec}}

We are now ready to evaluate the integrals in Eq.~(\ref{longlastingcase}). We consider the three approximations for the unequal time correlator of the 
anisotropic stress tensor mentioned above, namely incoherent, coherent and 
top hat. All we need is the anisotropic stress power spectrum taken at equal 
times, which is given in Eqs.~(\ref{Icalturb}) and (\ref{Icalmag}). These are 
then inserted into Eq.~(\ref{longlastingcase}). The GW power spectra obtained
in this way are always positive. The figures of the GW energy density spectra 
shown below are calculated for the set of parameter values given at the end of the  
previous section. Under the equipartition hypothesis, we fix the magnetic 
field energy density to the same value of the turbulent one: 
$\omt=\frac{\Om_B^*}{\Om_{\rm rad}^*}=\frac{2}{9}$\,. 

Even though we have demonstrated in Sec.~\ref{sec:howlong} that
MHD turbulence can last for many Hubble times, at the end of this section we 
also consider the case of GW generated by MHD turbulence confined in 
time to the duration of the phase transition. We find this analysis 
illuminating to understand our results and useful to compare with the 
results obtained in previous works. 

From now on we use a common notation for both the GW spectra generated by turbulence and by the magnetic field, with $s=v$ or $b$ denoting respectively the turbulence and magnetic field source. 

\vspace*{0.5cm}
\noindent
$\bullet$~{\bf  Incoherent approximation}: in this case the anisotropic stress 
spectrum at unequal times is given by (\cf Eq.~(\ref{inco1}) and 
(\ref{Icalturb}) resp (\ref{Icalmag})):
\be
\Pi(K_*,y,z)=\frac{9}{2}\,\pi\,\mathcal{C}_s^2\left(\frac{\Omega_S}{\Omega_{\rm rad}}(y)\right)^2L^3(y) \,\mathcal{I}_s(K_*,y,y)\,\,\delta(y-z)\,,
\ee
where we have chosen $\tau_L$ as the short characteristic time over which the source remains coherent. From the anisotropic stress formula given above we find the spectrum:
\bea
\left. \frac{d\Omega_{GW}h_0^2}{d\log k}\right|_{0}&=&12(2\pi)^2 \,
\mathcal{C}_s^2 \, \Om_{\rm rad,0}h_0^2 \left(\frac{g_0}{g_{\rm fin}}\right)^{\frac{1}{3}} \left(\oms\right)^2 K_*^3 \\
&\times&\left\{ \int_0^1 \frac{dy\,\,y^{3\gamma+2}}{\big[ y + \frac{\tin}{\tau_L} \big]^2} \,\,\II_s(K_*,y,y)
+\int_1^{y_{\rm fin}} \frac{dy\,\,y^{-7\gamma}}{\big[ y + \frac{\tin}{\tau_L} \big]^2} \,\,\II_s(K_*,y,y) \nonumber
\right\}\,
\eea
where $\tau_L=L_*/(2v_L)$ is the initial eddy turnover time at the scale $L_*$ 
for $y=1$, and $v_L=\sqrt{6\pi\mathcal{C}_v \omt}$ denotes the initial eddy 
turn-over speed, Eq.~(\ref{vLsquared}); moreover, see Sec.~\ref{sec:freedecay} 
\be
\frac{\tin}{\tau_L}=\frac{t_*}{\tau_L}-1~~~~~\mbox{where}~~~~~\frac{t_*}{\tau_L}=\frac{v_L}{v_b}\frac{\beta}{\mathcal{H}_*}\,,
\ee
$y_{\rm fin}$ is given by $(\tfin(k)-\tin)/\tau_L$ and $\tfin(k)$ is defined in 
Section \ref{sec:etafin}. The spectra $\II_s(K_*,y,y)$ are given in 
Eqs.~(\ref{Icalturb}) and (\ref{Icalmag}).

\begin{figure}[htb!]
\begin{center}
\includegraphics[width=8cm]{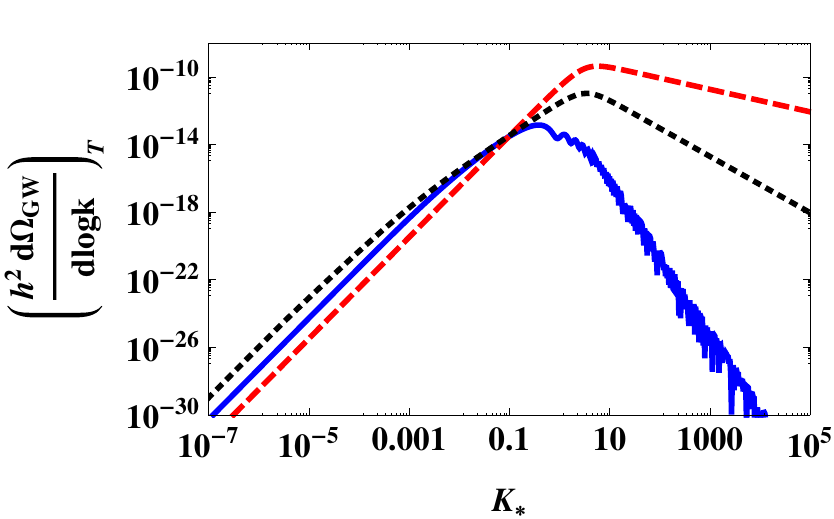}
\includegraphics[width=8cm]{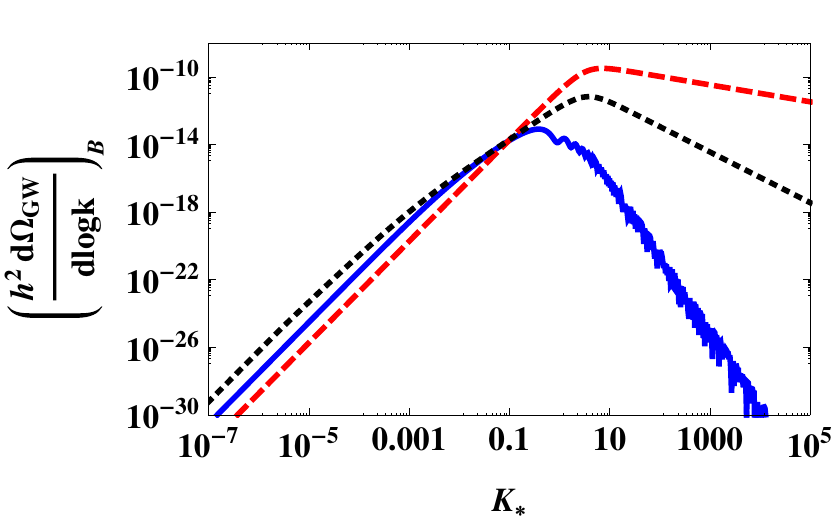}
\caption{\small The GW energy density spectrum in the incoherent (red, long-dashed), top hat (black, short-dashed) and coherent (blue solid) cases. Left from turbulence, right from magnetic field, for $T_*=100$ GeV, $\beta/\mathcal{H}_*=100$, $\oms=2/9$, $v_b=0.87$, $\gamma=2/7$, $g_{\rm fin}=47.75$ (and $x_c=1$ for the tophat case). \label{figall}}
\end{center}
\end{figure}

We recover the same behaviour as in section~\ref{sec:constant}. The GW power spectrum is proportional to the phase space volume $K_*^3$ times the source power spectrum. The slope at large scales is therefore $K_*^3$, the one at small scales is $K_*^{3+n}$ where $n=-11/3$ for the turbulence and $n=-7/2$ for the magnetic field. The peak is at $K_*^{\rm peak}\simeq 5.9$ for turbulence and $K_*^{\rm peak}\simeq 7$ for the magnetic field. The results are shown in Fig.~\ref{figall} for the usual choice of the parameter values. The incoherent approximation is the one leading to the highest peak amplitude:\bea
&& \left. \frac{d\Omega_{GW}h_0^2}{d\log k}\right|_{K_*^{\rm peak}}^{({\rm turb})}
 \simeq 4\times 10^{-10} \, \qquad K_*^{\rm peak} \simeq 5.9 \,, \\
&& \left. \frac{d\Omega_{GW}h_0^2}{d\log k}\right|_{K_*^{\rm peak}}^{({\rm mag})} 
\simeq 3\times 10^{-10} \, \qquad K_*^{\rm peak} \simeq 7 \,. 
\eea

\vspace*{0.5cm}
\noindent
$\bullet$~{\bf  Coherent approximation}: According to 
Eq.~(\ref{co1}), we have
\be
\Pi(K_*,y,z)=\frac{9}{2} \,\pi\, \mathcal{C}_s^2\,
\frac{\Omega_S}{\Omega_{\rm rad}}(y) L^{\frac{3}{2}}(y) \,
\frac{\Omega_S}{\Omega_{\rm rad}}(z) L^{\frac{3}{2}}(z)\,
\sqrt{\mathcal{I}_s(K_*,y,y)}\sqrt{\mathcal{I}_s(K_*,z,z)}\,.
\ee
The GW spectrum is positive by construction, but it is oscillatory:
\bea
\label{GWcoherent}
\left. \frac{d\Omega_{GW}h_0^2}{d\log k}\right|_{0}&=&12(2\pi)^2 \,\mathcal{C}_s^2\, \Om_{\rm rad,0}h_0^2 \left(\frac{g_0}{g_{\rm fin}}\right)^{\frac{1}{3}}
 \left(\oms\right)^2 K_*^3 \\
&\times & \left\{ \left[ \int_0^1 dy\, \frac{y^{3\gamma/2+1}}{ y + \frac{\tin}{\tau_L}} \,\sqrt{\II_s(K_*,y,y)}\cos \left(\frac{\pi K_*}{v_L} y \right) \right. \right. \nonumber\\
&+& \left. \int_1^{y_{\rm fin}} dy\,\frac{y^{-7\gamma/2}}{ y + \frac{\tin}{\tau_L}} \,\sqrt{\II_s(K_*,y,y)}\cos\left( \frac{\pi K_*}{v_L} y\right) \right]^2 \nonumber\\
&+& \left[ \int_0^1 dy\,\frac{y^{3\gamma/2+1}}{ y + \frac{\tin}{\tau_L}} \,\sqrt{\II_s(K_*,y,y)}\sin\left( \frac{\pi K_*}{v_L} y\right)\right.  \nonumber\\
&+& \left. \left. \int_1^{y_{\rm fin}} dy\,\frac{y^{-7\gamma/2}}{ y + \frac{\tin}{\tau_L}} \,\sqrt{\II_s(K_*,y,y)}\sin\left( \frac{\pi K_*}{v_L} y \right) \right]^2 \right\}\,. \nonumber 
\eea
As discussed in Ref.~\cite{thomas} and in Section \ref{sec:constant}, the calculation in the coherent case is more involved than the incoherent one, since the GW power spectrum is not simply proportional to the source power spectrum, but it is given by the square of its time Fourier transform. The source is characterized by the space correlation scale $L_*$ and the time correlation scale $\tau_L$, related by $L_*=2 v_L \tau_L$. Since $v_L\lesssim 0.4 $ (the upper bound is given by Eq.~(\ref{vL}) with $\vev{v^2}=1/3$), we have $L_* < \tau_L$. On scales larger than both the characteristic spatial correlation scale $L_*$ and time correlation scale $\tau_L$, the Fourier transform of the source is constant because the source is not correlated (white noise). Therefore, for wave-numbers $k\lesssim 2\pi/\tau_L< 2\pi/L_*$, corresponding to $K_*\lesssim L_*/\tau_L $, we recover the $K_*^3$ behaviour like in the incoherent case. However, for $k\gtrsim 2\pi/\tau_L$ the time Fourier transform is no longer constant and starts to decay as a power law, the exponent depending on the time differentiability properties of the source \cite{thomas}. Since we have chosen a source which is continuous but not differentiable at initial time $\tin$ (see the time evolution laws of the energy and the 
correlation scales given in Eqs.~(\ref{e:Leta}, \ref{e:OmTeta})), this implies 
a decay like $k^{-2}$ for the time Fourier transform of the source. Therefore, the GW power spectrum (\ref{GWcoherent}) (the square of the Fourier transform, multiplied by $K_*^3$) decays like $K_*^{-1}$ at intermediate scales $L_*/\tau_L \lesssim K_* \lesssim 1$. This behaviour is satisfied up to the wave-number corresponding to the spatial correlation scale: $k\simeq 2\pi/L_*$. Afterwards, the spectrum decays with a power law given by the time Fourier transform multiplied by the power law decay of the source. In the case of turbulence, for $K_*\gtrsim 1$ one has therefore the power law decay $K_*^{-14/3}=K_*^3\times (K_*^{-2})^2 \times (K_*^{-11/6})^2$; while in the magnetic field case this becomes $K_*^{-9/2}=K_*^3\times (K_*^{-2})^2 \times (K_*^{-7/4})^2$. 

\begin{figure}[htb!]
\begin{center}
\includegraphics[width=10cm]{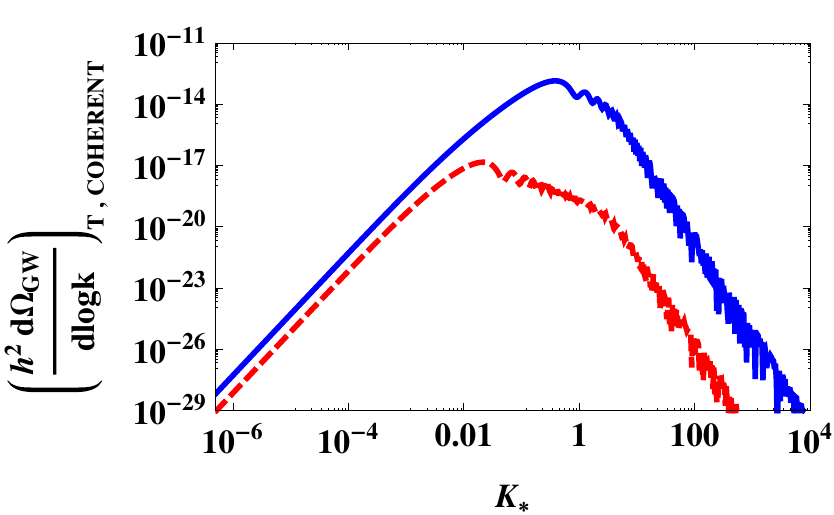}
\caption{\small The GW energy density power spectrum from turbulence in the coherent case for $T_*=100$ GeV, $\beta/{\cal H}_*=100$, $v_b=0.87$, $\gamma=2/7$, and $g_{\rm fin}=47.75$. Blue, solid:  $\Omega_{T*}/\Omega_{\mbox{\tiny rad}*}=2/9$. The peak is at about $K_*^{\rm peak}\simeq 0.4$, not far from the value ${L_*}/{\tau_L} = 2 (6\pi \mathcal{C}_v \Omega_{T*}/\Omega_{\mbox{\tiny rad}*})^{1/2}\simeq 0.8$. The small scale behaviour for $K_* > K_*^{\rm peak}$ is $K_*^{-14/3}$. Red, dashed: $\Omega_{T*}/\Omega_{\mbox{\tiny rad}*}=2/3\times 10^{-3}$. The peak position is at $K_*\simeq 0.02$, again not far from the value ${L_*}/{\tau_L} = 2 (6\pi \mathcal{C}_v \Omega_{T*}/\Omega_{\mbox{\tiny rad}*})^{1/2}\simeq 
 0.04$. In this case the slope $K_*^{-1}$ for intermediate wave-numbers is well visible. \label{figco}}
\end{center}
\end{figure}

The form of the power spectrum, in particular the wave-number at which the spectrum peaks is determined by the ratio $\frac{L_*}{\tau_L} = 2 \sqrt{6\pi \mathcal{C}_v\omt}$. In Fig.~\ref{figco}, we show the GW spectrum from turbulence for the coherent case for two values of this ratio: the case when the kinetic energy density in the turbulence in maximal $\vev{v^2}=1/3$, corresponding to $\omt=2/9$, and another one in which the kinetic energy involved is much smaller, $\vev{v^2}=10^{-3}$, corresponding to $\omt=2/3\times 10^{-3}$. Notice that not 
only the amplitude but also the peak position differs, and that the power law behaviour derived above is recovered. The form of the GW spectrum in the coherent case is also shown in Fig.~\ref{figall} for the usual choice of the parameters. The coherent approximation leads to the smallest peak amplitude:
\bea
%\label{e:amppeak.co}
&& \left. \frac{d\Omega_{GW}h_0^2}{d\log k}\right|_{K_*^{\rm peak}}^{({\rm turb})} \simeq 10^{-13} \,
 \qquad K_*^{\rm peak} \simeq 0.4 \,, \\
&& \left. \frac{d\Omega_{GW}h_0^2}{d\log k}\right|_{K_*^{\rm peak}}^{({\rm mag})} \simeq 8 \times 10^{-14} \,
 \qquad K_*^{\rm peak} \simeq 0.4 \,. 
\eea


\begin{figure}[htb!]
\begin{center}
\includegraphics[width=10cm]{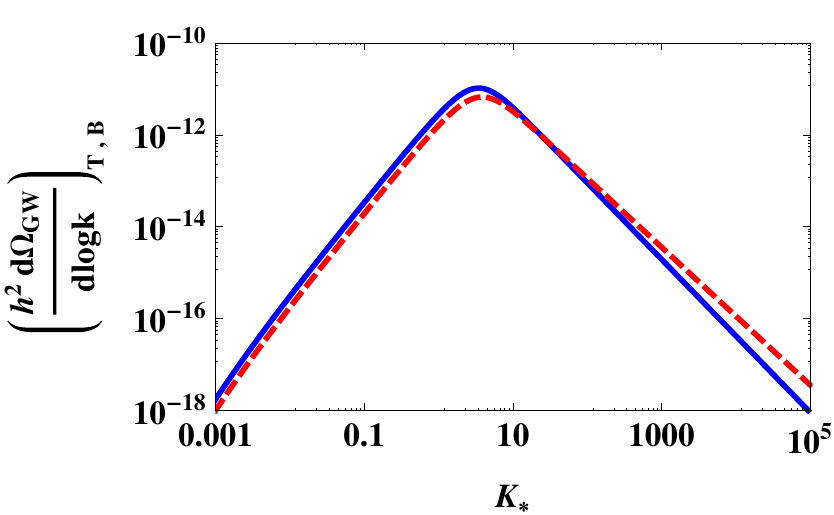}
\caption{\small The GW energy density spectrum in the top hat case. Blue, solid 
from turbulence and red, dashed from the magnetic field for $T_*=100$ GeV, $\beta/\mathcal{H}_*=100$, $\oms=2/9$, $v_b=0.87$, 
$\gamma=2/7$, $g_{\rm fin}=47.75$, $x_c=1$. \label{figtop}}
\end{center}
\end{figure}

\vspace*{0.5cm}
\noindent
$\bullet$~{\bf  Top hat approximation}: this is the most realistic case for the MHD turbulent source, since it mimics a de-correlation in time of the Kraichnan type as discussed in Section \ref{sec:anisotropic}:  
\bea
\Pi(K_*,y,z)&=&\frac{9}{4}\,\pi\,\mathcal{C}_s^2\left[ \left(\frac{\Omega_S}{\Omega_{\rm rad}}(y)\right)^2 L^{3}(y)\, \mathcal{I}_s(K_*,y,y) \,\Theta(z-y)\,\Theta\left(\frac{v_L x_c}{\pi K_*}-(z-y)\right) \right. \nonumber \\
&+&\left. \left(\frac{\Omega_S}{\Omega_{\rm rad}}(z)\right)^2 L^{3}(z)\, \mathcal{I}_s(K_*,z,z)\, \Theta(y-z)\,\Theta\left(\frac{v_L x_c}{\pi K_*}-(y-z)\right) \right]\,.
\eea
We choose the value $x_c=1$, as in the Kraichnan model, so that the integral determining the GW spectrum is positive. It is given by:
\bea
\nonumber
\left. \frac{d\Omega_{GW}h_0^2}{d\log k}\right|_{0}&=&12 (2\pi)^2 \,\mathcal{C}_s^2\, \Om_{\rm rad,0}h_0^2\left(\frac{g_0}{g_{\rm fin}}\right)^{\frac{1}{3}}
 \,\left(\oms \right)^2 \, K_*^3 \\
&\times &  \left[ \int_0^1 dy\, \frac{y^{3\gamma+2}}{ y+ \frac{\tin}{\tau_L}} \,\II_s(K_*,y,y) \int_y^{y_{\rm top}}  \frac{dz}{ z+ \frac{\tin}{\tau_L}} \cos\left( \frac{\pi K_*}{v_L} (z-y)\right) \nonumber \right.\\
&+& \left. \int_1^{y_{\rm fin}}dy\, \frac{y^{-7\gamma}}{ y + \frac{\tin}{\tau_L}} \,\II_s(K_*,y,y) \int_y^{y_{\rm top}}  \frac{dz}{ z + \frac{\tin}{\tau_L}} \cos\left( \frac{\pi K_*}{v_L} (z-y)\right)  \right] \,,
\eea 
where 
\be
y_{\rm top}={\rm min}\left[ y_{\rm fin}~,~y+\frac{v_L\,x_c}{\pi\,K_*} \right]\,.
\ee
As in the incoherent case, the spectrum bears no relation with the time Fourier transform of the source (\cf \cite{thomas}). The integral in $z$ can be estimated simply as the integrand evaluated at the lower bound, multiplied by one oscillation period: therefore, at high wave-numbers we expect the slope $K_*^{-5/3}$ for the turbulent case, and $K_*^{-3/2}$ for the magnetic field case. Moreover, the peak position corresponds to the spatial correlation scale of the source: $K_*\simeq 3.5$ for the turbulence and $K_*\simeq 3.7$ for the magnetic field. The result is shown in Fig.~\ref{figall} and \ref{figtop}, for $x_c=1$; the dependence on the value of $0<x_c<\pi$ is weak. In this case the amplitude at the peak takes an intermediate value between the incoherent and coherent cases: 
\bea
%\label{e:amppeak.top}
&& \left. \frac{d\Omega_{GW}h_0^2}{d\log k}\right|_{K_*^{\rm peak}}^{({\rm turb})} \simeq 10^{-11} \,
 \qquad K_*^{\rm peak} \simeq 3.5 \,, \\
&& \left. \frac{d\Omega_{GW}h_0^2}{d\log k}\right|_{K_*^{\rm peak}}^{({\rm mag})} \simeq 7\times10^{-12} \,
 \qquad K_*^{\rm peak} \simeq 3.7 \,. 
\eea

\section{Discussion  \label{s:dis}}

\subsection{Comparison with the gravitational wave spectrum from bubble collisions}

In the case of bubble collisions~\cite{Caprini:2007xq}, we have also studied
the different assumptions for the unequal time power spectrum of the source: 
coherent, incoherent and top-hat. 
Numerical simulations of bubble collisions \cite{Huber:2008hg} indicate that 
the coherent case is the relevant one for bubbles. This can be understood 
by the following argument: we consider a bubble collision event starting at 
time $t_n$ with tensor anisotropic stress given by $f_n(\bk,t-t_n)$. Summing
over all the collision events of the phase transition we obtain for the
total anisotropic stress of bubble collisions (\cf \cite{thomas})
\bea
 \langle\Pi(\bk,t)\Pi^*(\bk',t')\rangle =  
 \sum_{n=1}^N\sum_{m=1}^N \langle e^{i(\bk\cd\bx_n-\bk'\cd\bx_m)} 
\hat f_n(\bk,t-t_n)\hat f_m^*(\bk',t'-t_m)\rangle~,
\eea
where $\bx_n$ is the center of the $n$-th collision event. To recover the simulation result, we now have to make two fundamental assumptions.  
First we assume that different collision events are not correlated,
\be
\langle e^{i(\bk\cd\bx_n-\bk'\cd\bx_m)}\rangle 
=  V^{-1}\de_{nm}
   \delta(\bk-\bk') \,, \label{centres} 
\ee
here the volume $V$ restores the dimensions. This leads to
\be\label{e:Pibub}
\Pi(k,t,t') = \frac{1}{(2\pi)^3V}\sum_{n=1}^N 
\langle f_n(\bk,t-t_n) f_n^*(\bk,t'-t_n)\rangle \,.
\ee
Second, we assume that one typical collision event is totally coherent, so that
$\langle f_n(\bk,\De t) f_n^*(\bk,\De t')\rangle = f(\bk,\De t)
 f^*(\bk, \De t')$ where $f(\bk,\De t) \equiv 
\sqrt{\langle |f_n(\bk,\De t)|^2\rangle}$ is the square root of the power 
spectrum of a typical collision event, and $\De t=t-t_n$. Furthermore, since 
bubbles only exist during the phase transition the duration of which is much 
shorter than one Hubble time, the expansion of the universe can be neglected 
and the anisotropic stress spectrum is a function  of the time difference
$t-t'=\De t-\De t'$ only. Equation (\ref{e:Pibub}) becomes
\be\label{e:Pibub2}
\Pi(k,t,t') = \frac{2N}{(2\pi)^3V}
f(\bk,\De t)f(\bk,\De t') = \sqrt{\Pi(k,t,t)\Pi(k,t',t')}\,,
\ee
which is the form of the coherent approximation. 
Therefore, assuming that one collision event is totally coherent in time and 
that different collisions are uncorrelated implies that bubble collisions 
represent a totally coherent source of GWs.

However, for turbulence, we do not expect this to hold. There are no well 
isolated uncorrelated events which can be treated independently. 
Besides, one expects correlations to decay in time over a timescale which
is related to the spatial extension of the source. This has motivated
the Kraichnan de-correlation ansatz given in \cite{Kraichnan} and Eq.~(\ref{Krai}), which we have simplified to the the top hat de-correlation in 
Section~\ref{sec:anisotropic} in order to obtain a positive kernel, see 
Eq.~(\ref{tophatpi}) and the discussion  following it. To summarize,  we can conclude that bubble collisions are well represented by the coherent case, while MHD turbulence is well represented by the top-hat case.

\subsection{Comparison with short-lasting turbulence}

Once MHD turbulence is generated, it decays following the `absolute' time 
dependence described in Section~\ref{sec:turbulence}, which applies to freely 
decaying, non-helical turbulence. In this section we compare our results
with the ones from MHD turbulence that lasts for less than one Hubble time 
and where  the time-dependent decay is neglected. This was always assumed 
in previous analyses. This allows us to verify the general statements of 
Section~\ref{sec:constant} in a more realistic case. Note however that, with 
respect to previous analyses which assumed either a discontinuous~\cite{CDturb}
or a stationary \cite{turb1,Tinaturb,Kahniashvili:2008pe} source, here we 
consider a source with a finite time duration, but continuous in time. The 
importance of having a continuous source has been discussed in~\cite{thomas}.  

The aim of this section is simply to test the results of section 
\ref{sec:constant} in a realistic case.
We therefore concentrate only on the source from the turbulent velocity 
field, and we do not discuss the magnetic field for which the results are similar. 
We set the duration of the source to one eddy turnover time $\tau_L$: 
$\tfin=\tin+\tau_L$ \cite{turb1}. We model the switching on and off of the 
source with the same function as in bubble collisions~\cite{thomas}:
\be
f(y)=4y(1-y)
\ee
with $y=(t-\tin)/\tau_L$. Since $L$ and $\Omega_T/\Omega_{\rm rad}$ do not evolve in time, the turbulent spectrum becomes 
\be
P_v(K_*)=\frac{3}{2}\,\mathcal{C}_v\,\omt\,L_*^3\,
  \frac{K_*^2}{[1+K_*^2]^{17/6}} f(y)\,,
\ee
and the anisotropic stress power spectrum is
\be
\Pi(K_*,y,y)=\frac{9}{2}\, \pi \,\mathcal{C}_v^2 \left(\omt\right)^2L_*^3\,\,
  \mathcal{I}_v(K_*)\,f^2(y)\,, 
\ee
where $\mathcal{I}_v(K_*)$ is given by Eq.~(\ref{Icalturb}) with $y=1$. 
Inserting this in Eq.~(\ref{shortla}) one finds after some manipulations 
\bea
\left. \frac{d\Omega_{GW}h_0^2}{d\log k}\right|_{0}=12 (2\pi)^2 \,\mathcal{C}_v^2
  \Om_{\rm rad,0}h_0^2 \left(\frac{g_0}{g_{\rm *}}\right)^{\frac{1}{3}} 
\left(\omt\right)^2 \,\,(\mathcal{H}_*\tau_L)^2 K_*^3\,\,\mathcal{I}(K_*)\times
 \nonumber \\
\hspace*{4cm} \int_0^1dy\int_0^1dz\cos\left(\frac{\pi\,K_*}{v_L}(y-z)\right)
F(y,z) 
\eea
with
\be
F(y,z) =
\left\{\begin{array}{ll}
f^2(y)\,\delta(y-z) & \mbox{incoherent} \\
f(y)\,f(z) & \mbox{coherent} \\
\frac{1}{2}\left[ f^2(y)\Theta(z-y)\,\Theta\left(\frac{v_L\,x_c}{\pi\,K_*}
 -(z-y)\right) + y\leftrightarrow z)\,\right]   & \mbox{top~hat.}
\end{array}\right.
\ee

\begin{figure}[htb!]
\begin{center}
\includegraphics[width=10cm]{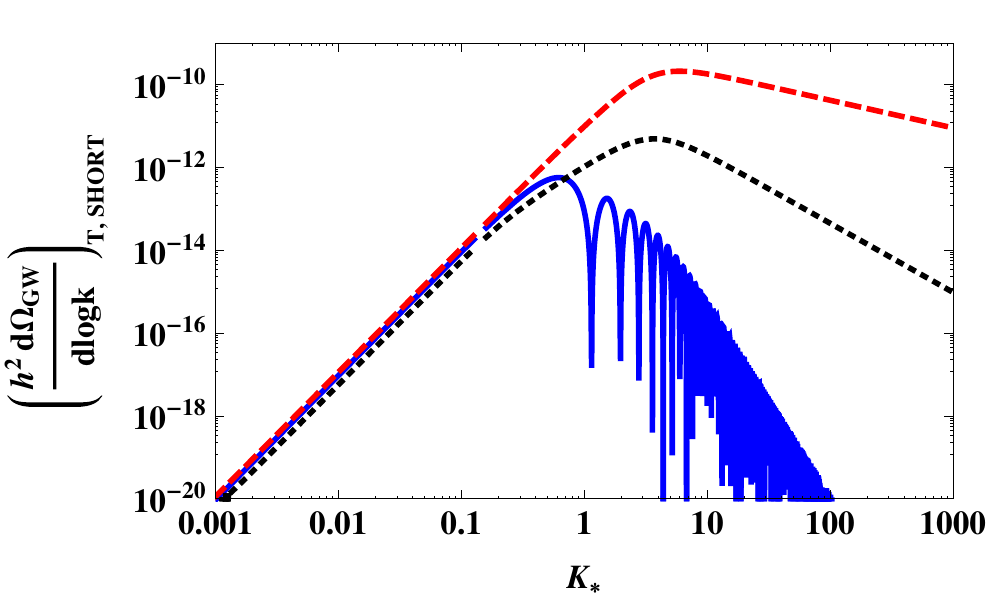}
\caption{\label{figshortall} \small The GW energy density spectrum for short lasting turbulence. Blue, solid: coherent; black, short-dashed: top hat; red, long-dashed: incoherent.}
\end{center}
\end{figure}

\begin{figure}[htb!]
\begin{center}
\includegraphics[width=10cm]{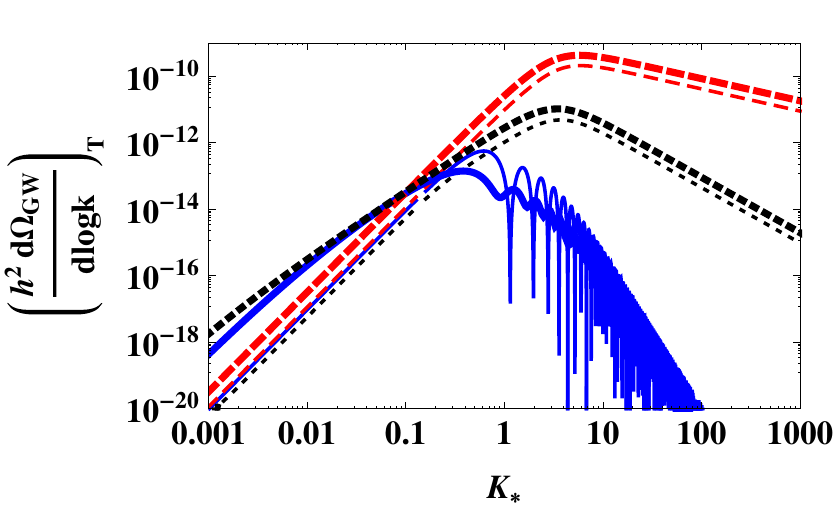}
\caption{\label{allcomparison} \small Comparison between the long-lasting case, represented by thick lines and 
the short-lasting one, represented by thin lines (incoherent: red, long-dashed; coherent: blue, solid; top hat: black, short-dashed).}
\end{center}
\end{figure}
Fig.~\ref{figshortall} shows the GW energy density spectra from short lasting 
turbulence in the incoherent, coherent and top hat cases. The 
relative amplitudes between the different approximations is similar to the long 
lasting case and the peak positions are also not very different. In 
Fig.~\ref{allcomparison} we compare the long-lasting and the short-lasting
cases. 

One first notices that the change in amplitude of the GW spectrum is 
generically much less than expected from our general arguments in 
Section~\ref{sec:constant}. This is because our 
realistic source, unlike the one considered in 
Section~\ref{sec:constant}, is decaying with the decay time $\tau_L$, which here corresponds to the  
duration of the short lasting source. Therefore, the fact 
that the source is long lasting does not amplify the signal as expected, because
 its characteristic decay time is the same as the duration of the 
short lasting source. In the incoherent approximation, the overall amplification of the 
long lasting source is about a factor two, instead of the factor 
$(\tau_L\HH_*)^{-1}$ which we found for the toy model. The long lasting 
coherent approximation, on the other hand, is amplified at large scales by about a 
factor of $(\tau_L\HH_*)^{-1}\simeq 80$ compared to the short lasting one, 
instead of the expected $(\tau_L\HH_*)^{-2}$. 
As already seen in 
Section~\ref{sec:constant}, in the coherent case the signal at the peak from 
short-lasting turbulence is higher than the long-lasting result: this comes 
from the fact that in the long-lasting case interference can reduce the final
amplitude if the source is present and coherent over several oscillation 
periods. 

The top-hat case, which we did not analyze in Section~\ref{sec:constant}, 
shows an intermediate behavior between the incoherent and coherent ones:  
at long wavelengths, $K_*\ll K_*^{\rm peak}$, the long lasting signal is 
enhanced by about two orders of magnitude, \ie the same amplification as for the 
coherent case, $(\tau_L\HH_*)^{-1}$; around the peak, the long and short 
lasting top hat cases differ by about a factor two, equivalent to the 
incoherent case. 

To summarize, due to the fact that the long lasting source decays with a 
characteristic time $\tau_L$ corresponding to the duration of the short 
lasting source, the amplification factor $(\tau_L\HH_*)^{-2}$, expected at large 
scales, is reduced to $(\tau_L\HH_*)^{-1}$ for the coherent and top-hat approximations, 
and the amplification is virtually absent in the incoherent approximation.
In all cases it is a factor $(\tau_L\HH_*)$ less than expected due to the
rapid decay of the source.

\section{Conclusion \label{s:con}}
In this work, we have calculated the GW emission from MHD 
turbulence generated during a first order phase transition and freely 
decaying afterwards. This is the first paper which takes into account the free 
decay of turbulence, and models the source in a continuous fashion. 

For the source power spectrum we use a new ansatz that interpolates 
analytically between the large scale and small scale behaviors, determined 
respectively by causality and by the Kolmogorov (or Iroshnikov-Kraichnan) 
theory. Previous analyses had either considered only the Kolmogorov range, 
ignoring the large scale part of the spectrum and continuing the Kolmogorov 
slope up to the peak \cite{turb1,turb2,Tinaturb,Kahniashvili:2008pe}, or joined
 the two behaviors at the peak \cite{CDturb}. This caused an overestimation 
of the source spectrum amplitude at the peak of about a factor six, leading to 
an overestimation of nearly two orders of magnitude for the GW spectrum (\cf for example Eq.~(56) of \cite{CDturb}). The 
interpolating formula which we adopt here models the spectrum of MHD turbulence
in a more realistic way. 

We also take into account the time de-correlation of MHD turbulence following the model proposed by Kraichnan in \cite{Kraichnan}. However, in order to recover a positive result for the GW energy density spectrum, we cannot directly apply the Kraichnan de-correlation in the velocity and magnetic field power spectra, but we have to model it in Fourier space as a de-correlation of the anisotropic stress power spectrum. We claim that in the case of MHD turbulence, neither the coherent \cite{CDturb} nor the stationary \cite{turb1,Tinaturb,Kahniashvili:2008pe} approximations previously used in the literature are the correct ones, but the anisotropic stresses at different times have to be modeled in a way similar to our top hat ansatz.

Moreover, previous analyses have considered either a MHD source which is discontinuous in time (\ie instantaneous turning on of the Kolmogorov spectrum) \cite{CDturb} or a source which is stationary (\ie neglecting the fact that the source is actually turned on and off) \cite{turb1,Tinaturb,Kahniashvili:2008pe}. Here instead the source is continuous in time, starting with zero energy and building up the Kolmogorov spectrum after one eddy turnover time. It then starts free decay, since the stirring due to the phase transition lasts only for about one eddy turnover time. Due to the free decay, the source is absent on scales smaller than the time dependent  Kolmogorov microscale, and is completely dissipated once the Kolmogorov microscale has reached the energy injection scale and the entire Kolmogorov range has decayed. We have evaluated the temperature of the universe, $T_{\rm fin}$, at which this happens, as a function of the temperature at which  MHD turbulence is generated, $T_*$, and we have found that the process of dissipation lasts for many Hubble times. For instance, for the EW phase transition, $T_* \sim $ 100 GeV and $T_{\rm fin}\sim 120$ MeV. Therefore, MHD turbulence has to be modeled as a long lasting GW source.  
Nevertheless, since the characteristic decay time of the source is still given by the eddy turnover 
time which is much smaller than the expansion time of the Universe, the 
amplification due to the long duration is less significant than what is 
expected from a source which is not decaying in time. Especially the peak amplitude is enhanced 
only by about a factor of two due to the long duration of the source. On the other hand, the long duration of the source becomes
important on very large scales: here the GW spectrum from the long lasting 
source is amplified with respect to the short lasting case by about two orders 
of magnitude.

The top hat ansatz together with the time continuity of the source have some interesting consequences on the peak position and amplitude of the GW spectrum. In Ref.~\cite{thomas} it has been shown that, in the coherent case, time continuity affects the slope of the spectrum at small scales and also moves the peak of the GW spectrum from the characteristic length scale of the source (here $L_*$) to its characteristic time scale (here $\tau_L$). If the two scales are well separated (for instance if the fluid velocity is significantly smaller than the speed of light), this causes a reduction of the amplitude at the peak. The fact that for MHD turbulence the top hat ansatz is the relevant one, fixes the peak position of the GW spectrum at the characteristic length scale of the turbulent source. 

\begin{figure}[htb!]
\begin{center}
\includegraphics[width=8.1cm]{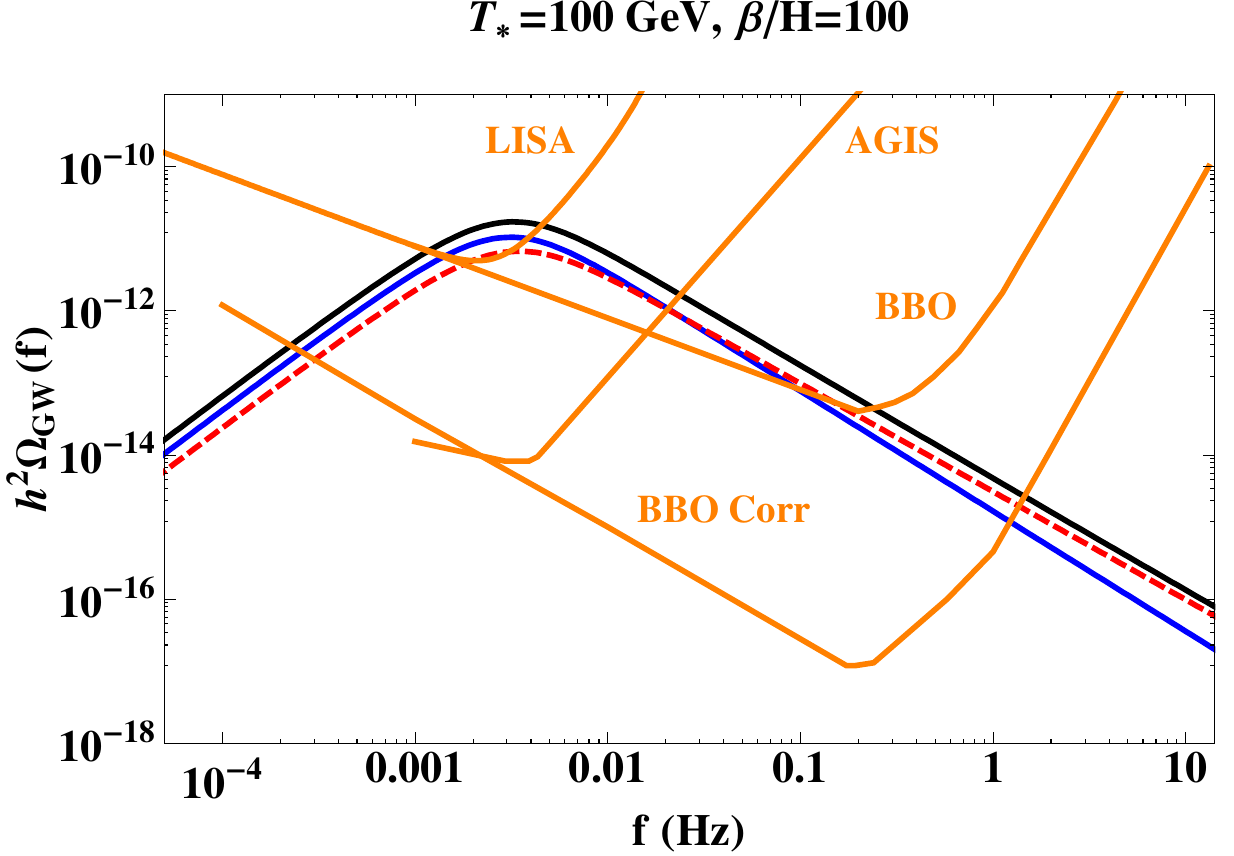}
\includegraphics[width=8.1cm]{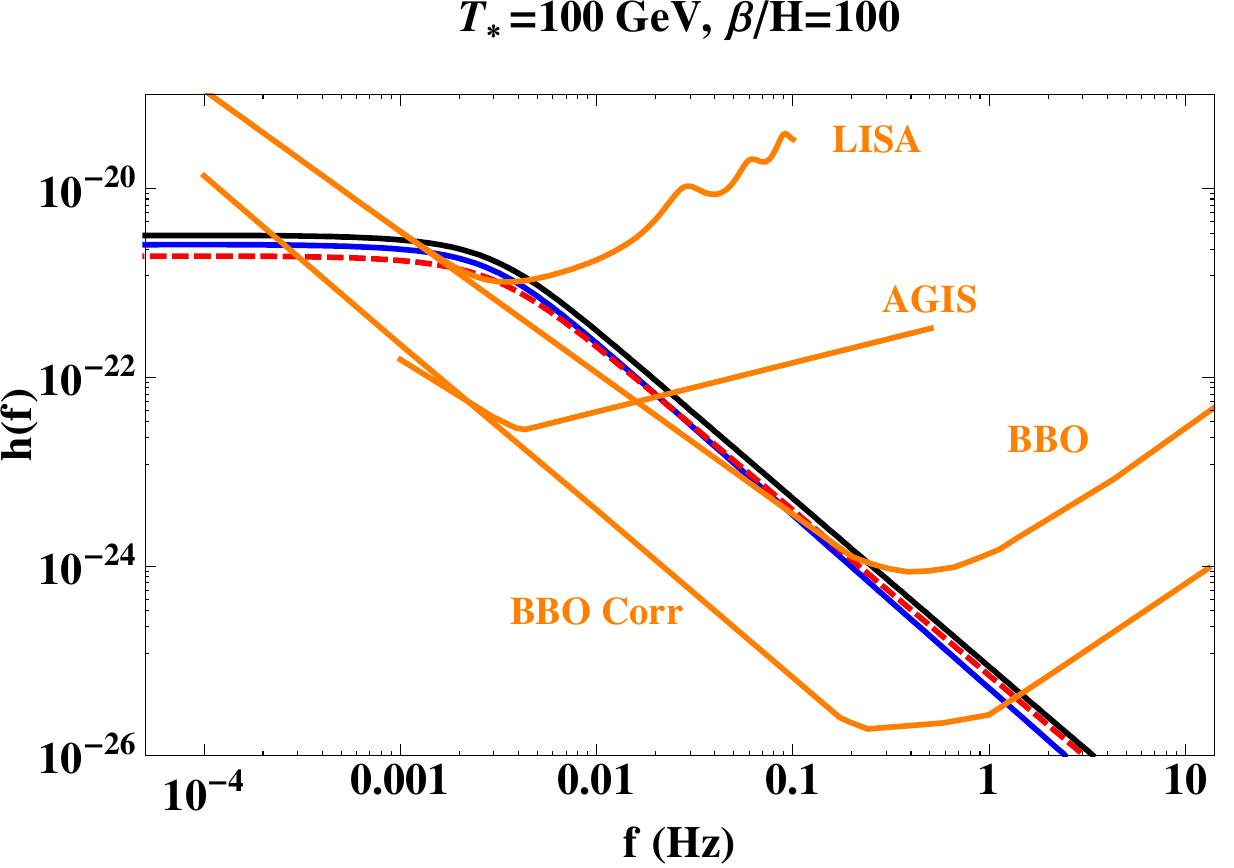}
\caption{\small The GW energy density (left) and the GW characteristic amplitude (right) from turbulence only (blue, solid), magnetic field only (red, dashed) and the total MHD turbulence (black, solid) generated at the EW phase transition with $T_*=100$ GeV, $\beta/\mathcal{H}_*=100$, $\Omega_{S*}/\Omega_{\mbox{\tiny rad}*}=2/9$, $v_b=0.87$, $\gamma=2/7$, $x_c=1$ and $g_{\rm fin}=47.75$, together with the sensitivities of LISA \cite{LISA}, BBO \cite{Harry:2006fi}, AGIS \cite{Dimopoulos:2008sv} and `BBO Corr' (improved from BBO with data analysis) taken from \cite{Buonanno:2004tp}. \label{fig:hf}}
\end{center}
\end{figure}

To summarize, our final result for the GW spectrum from MHD turbulence for the most realistic 
case, the top-hat de-correlation, has the following main features:
\begin{itemize}
\item The peak frequency is given by $K_*\simeq 3.5$ for the turbulence and $K_*\simeq 3.7$ for the magnetic field. Given the degree of precision of our analytical estimate, we can neglect this small difference. The peak frequency from MHD turbulence thus corresponds to 
\bea
f_{\rm peak}^{\rm MHD}&=&\frac{k_{\rm peak}}{2\pi}\simeq \frac{3.5}{L_*} = 3.5\,  \frac{\beta}{2v_b} \\
&\simeq&  3\times 10^{-2} \,{\rm mHz}\,\left( \frac{g_*}{100}\right)^{1/6}\frac{T_*}{100\,{\rm GeV}}\, \frac{\beta}{\mathcal{H}_*}\, \frac{1}{v_b}\nonumber \\
&\simeq& 3.4~{\rm mHz}\,,\nonumber
\label{peakfrequency}
\eea
where we have used $\mathcal{H}_*=(g_0/g_*)^{1/6}\,(T_*/T_0)\,H_0\sqrt{\Om_{\rm rad}}$, and the last line gives the value  
for the EW phase transition with $\beta/\HH_* =100$, $T_*=100$ GeV, $v_f=1/\sqrt{3}$ and $v_b\simeq 0.87$. 

\item The peak amplitude for the above values of the parameters is $h^2_0\Om_{GW}(K_*^{\rm peak})\simeq 
10^{-11}$. Using that $h(f)= 1.26\times 10^{-15}
\sqrt{h^2_0\Om_{GW}(f)}({\rm mHz}/f)$ (see \cite{GW1}), we obtain the maximal GW amplitude 
$h(f_{\rm peak}) \simeq 2 \times 10^{-21}$ which is detectable with LISA (see Fig.~\ref{fig:hf}).  
The suppression of the peak amplitude by more than one order of magnitude 
compared \eg to Ref.~\cite{Tinaturb},  
is mainly due to the fact that we use the more realistic interpolating spectrum 
for the MHD turbulent source (see Fig.~\ref{comparison}).

\item The slope of $d\Om_{GW}/d\log(k)$ is
the usual $k^3$ on large scales $k<k_{\rm peak}$, but on small scales, $k>k_{\rm peak}$, it decays only like $k^{-5/3}$ and $k^{-3/2}$ for turbulence and magnetic field 
respectively, see Fig.~\ref{figtop}.
\end{itemize}

\begin{figure}[htb!]
\begin{center}
\includegraphics[width=8.1cm]{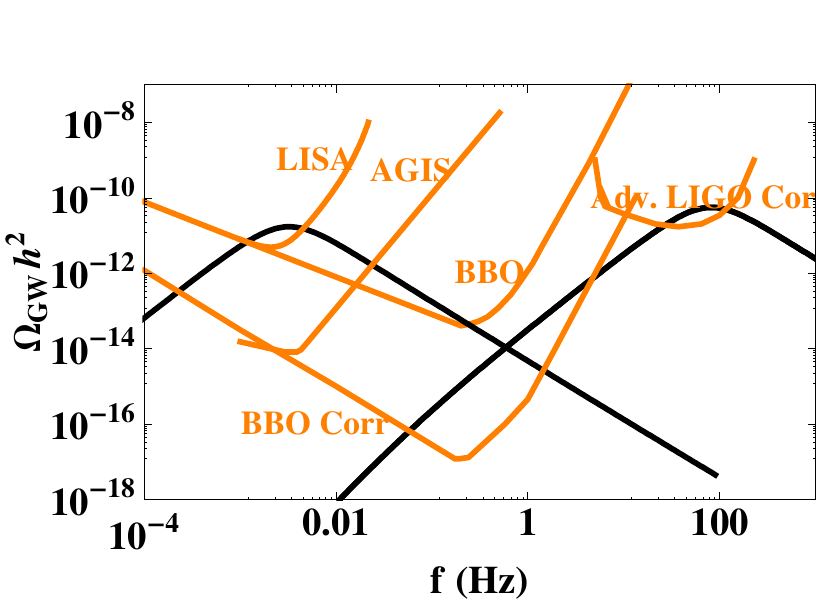}
\includegraphics[width=8.1cm]{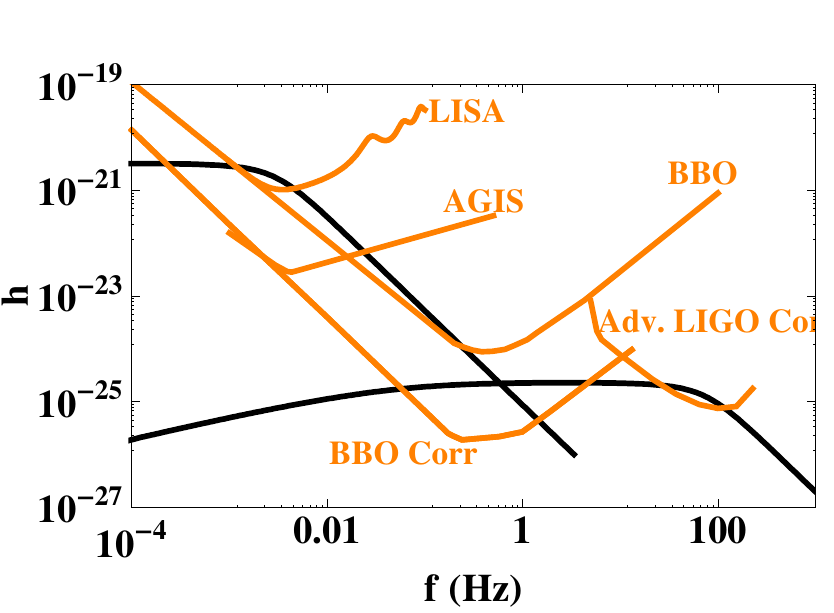}
\caption{\small Sensitivities of LISA, AGIS, BBO and Advanced LIGO (orange) compared with two GW spectra (black) generated by MHD turbulence  from  a  phase transition at  respectively $T_*=100$ GeV with $\beta/\mathcal{H}_*=100$, and $T_*=5.10^6$ GeV with $\beta/\mathcal{H}_*=50$; $\Omega_{S*}/\Omega_{\mbox{\tiny rad}*}=2/9$, $v_b=0.87$, $\gamma=2/7$, and $x_c=1$. The Advanced LIGO sensitivity is optimized 
by making use of correlations between two ground-based detectors \cite{Buonanno:2003th}. \label{fig:detection}}
\end{center}
\end{figure}

In Figs.~\ref{fig:hf} and \ref{fig:detection} we compare the GW spectrum from MHD turbulence with the experimental sensitivities of LISA \cite{LISA}, AGIS \cite{Dimopoulos:2008sv}, LIGO \cite{LIGO} and the Big Bang Observer (BBO)  \cite{Harry:2006fi}. Note that the sensitivity curve of LISA plotted in the figures represents only the noise of the instrument, and the data analysis can actually improve the detection down to a level of $h^2_0\Om_{GW}\sim 5\times 10^{-13}$ \cite{cornish}. 

With respect to the GW signal from bubble collisions analyzed in 
Ref.~\cite{Caprini:2007xq}, the peak frequency of the GW spectrum from MHD 
turbulence is larger by about a factor two. Note however that, as already pointed out in \cite{thomas}, the 
work of Ref.~\cite{Caprini:2007xq} has to be corrected in two aspects: first, 
the source analyzed there was not continuous, and secondly, the relevant 
approximation for the anisotropic stresses generated by bubble collisions is 
the coherent and not the top hat one. These modifications are work in progress 
and may well lead to some correction in the results of \cite{Caprini:2007xq}.
The peak amplitude of the GW signal from MHD turbulence is 
somewhat higher than the signal from bubble collisions, but it 
decays faster than the $k^{-1}$ decay which has been seen in the latest 
simulations of bubble collisions~\cite{Huber:2008hg}. 

Nevertheless, there is still 
considerable uncertainty in our analytical modeling which probably can only 
be addressed by numerical simulations of relativistic MHD turbulence of the 
kind developed after a first order phase transition. For example, in this 
work we have added the results from the turbulent velocity field and the 
magnetic field incoherently. One could argue, however, that these fields are 
correlated and have both a Kolmogorov spectrum in the inertial range. One 
could then use $s_i =b_i+v_i \simeq 2v_i$ as transverse vector-field with
$\Pi_{ij}({\bf s}) \simeq 4\Pi_{ij}({\bf v})$. Instead of the results presented 
here we would then obtain 16 times the GW energy density spectrum from 
turbulence, which would enhance the total result for $\Om_{GW}$ by about a 
factor of 8 and the one for $h(f)$ by $\sqrt{8}$.
Therefore, our results are probably to be taken within about a factor of 
a few for the GW amplitude $h(f)$ and within an order of magnitude 
for the energy density $d\Om_{GW}/d\log(k)$.


\section*{Acknowledgments}
We thank Kandu Subramanian for useful suggestions, and we acknowledge 
discussions with  Axel Brandenburg, Karsten Jedamzik and Tina Kahniashvili.
We are grateful for the hospitality of Nordita where this work was completed.
 RD acknowledges support from the Swiss National Science Foundation.
The work of GS is supported by the European Research Council. CC thanks the 
University of Geneva and CERN for hospitality.

\appendix

\section{Analytical expressions for Section~\ref{sec:constant}}
\label{Appen:analyt}

Here we give the full expression for Eqs.~(\ref{constantshort}) and (\ref{constantco}).

{\bf $\bullet$ Incoherent constant source}
\vspace*{0.5cm}
\noindent
\begin{eqnarray}
 F(\tin,\tfin,\Det) &=& 
\left\{\begin{array}{l}
\vspace*{0.3cm}
\big(\frac{g_0}{g_{\rm fin}}\big)^{\frac{1}{3}}\,8\Big[ 1-
  \frac{\tfin}{\Det}\log\!\big(\frac{\tfin}{\tfin-\Det/2}\big)
 -\frac{\tin}{\Det}\log\!\big(\frac{\tin+\Det/2}{\tin}\big)
  \Big] \\ 
\vspace*{0.3cm}
\simeq  
\big(\frac{g_0}{g_{\rm fin}}\big)^{\frac{1}{3}} \frac{\Det}{\tin}  \qquad    
   \qquad \quad\mbox{long-lasting,} \vspace*{0.1cm}\\
 \big(\frac{g_0}{g_*}\big)^{\frac{1}{3}}\,
 \frac{(2\pi)^2}{3}\,\big(\frac{\Det}{\tin}\big)^2 ~ \qquad\mbox{short-lasting.}
\end{array}\right.
\label{ap:constantshort}
\end{eqnarray}

\vspace*{0.5cm}
\noindent
{\bf $\bullet$ Coherent constant source}
\begin{eqnarray}
  F(\xin,\xfin,\De x) &=&\label{ap:factorcoherent}
\left\{\begin{array}{l}
\vspace*{0.3cm}
\big(\frac{g_0}{g_{\rm fin}}\big)^{\frac{1}{3}} 
\frac{4}{\De x^2}\left\{ \Big[\xfin \Ci(\xfin) + \xin \Ci(\xin)  \right. \nonumber \\ 
  \qquad    - (\xfin-\xt/2) \Ci(\xfin - \xt/2) - (\xin+\xt/2) \Ci(\xin + \xt/2) \nonumber \\
   \qquad  - \sin(\xfin) - \sin(\xin)  + \sin(\xfin - \xt/2)  + \sin(\xin + \xt/2)\Big]^2  \nonumber \\ 
   \qquad  + \Big[  \xfin \Si(\xfin) + \xin \Si(\xin) \nonumber \\ 
  \qquad  - (\xfin-\xt/2) \Si(\xfin - \xt/2)  - (\xin+\xt/2) \Si(\xin + \xt/2)  \nonumber \\
\vspace*{0.3cm}
 \qquad   + \left. \cos(\xfin) + \cos(\xin) - \cos(\xfin - \xt/2)  - \cos(\xin + \xt/2)\Big]^2 \right\} \nonumber \\
\vspace*{0.4cm}
\qquad \simeq \big(\frac{g_0}{g_{\rm fin}}\big)^{\frac{1}{3}} \, \left[
 \left(\Ci(\xfin) - \Ci(\xin)\right)^2 +  \left(\Si(\xfin) - 
    \Si(\xin)\right)^2\right] +{\cal O}(\xt) \\
\qquad \mbox{long-lasting\,,}\nonumber\\   \\
\big(\frac{g_0}{g_*}\big)^{\frac{1}{3}}\,
\frac{64(2\pi)^2}{\xin^2}\,
    \frac{\sin^4((\xfin-\xin)/4)}{(\xfin-\xin)^2}  \quad
 \mbox{short-lasting.}
\end{array}\right.
 \end{eqnarray}

\section{Viscosity and magnetic diffusivity}
\label{Appen:viscosity}

The Reynolds number defined in Eq.~(\ref{Reynold_number}) is inversely 
proportional to the kinematic viscosity $\nu$ given by
\be
\nu=\frac{\bar{\eta}}{\rho+p},
\ee
where 
$\bar{\eta}$ is the 
shear viscosity. The kinematic viscosity is the transport coefficient that 
characterizes the diffusion of transverse momentum due to collisions in a 
medium, and is roughly the mean free path $\ell_{\rm mfp}$ of excitations. 
In fact one has \cite{Weinberg:1971mx}
\be
\bar{\eta}=\frac{4}{15} \frac{\pi^2}{30} g_* T^4 \ \ell_{\rm mfp} \qquad \mbox{so~that}\qquad \nu=\frac{\ell_{\rm mfp}}{5}\,.
\ee
The largest viscosity comes from the weakest interactions, since it is inversely proportional to the scattering 
cross section of the processes responsible for transport. 
 Simple parametric estimates using kinetic theory show that 
the shear viscosity at high temperature (where $T$ is much larger 
than the mass of the diffusing particle) behaves as (to leading-log accuracy): 
\be
\bar{\eta}= C \frac{T^3}{g^4 \log g^{-1}} 
 \ee
where  $g$ is the appropriate coupling constant (depending on the temperature and the length scale at which one wants to compute the Reynolds number) and 
$C$ is a numerical coefficient that can only be obtained from a detailed analysis.

\begin{figure}[htb!]
\begin{center}
\includegraphics[width=10cm]{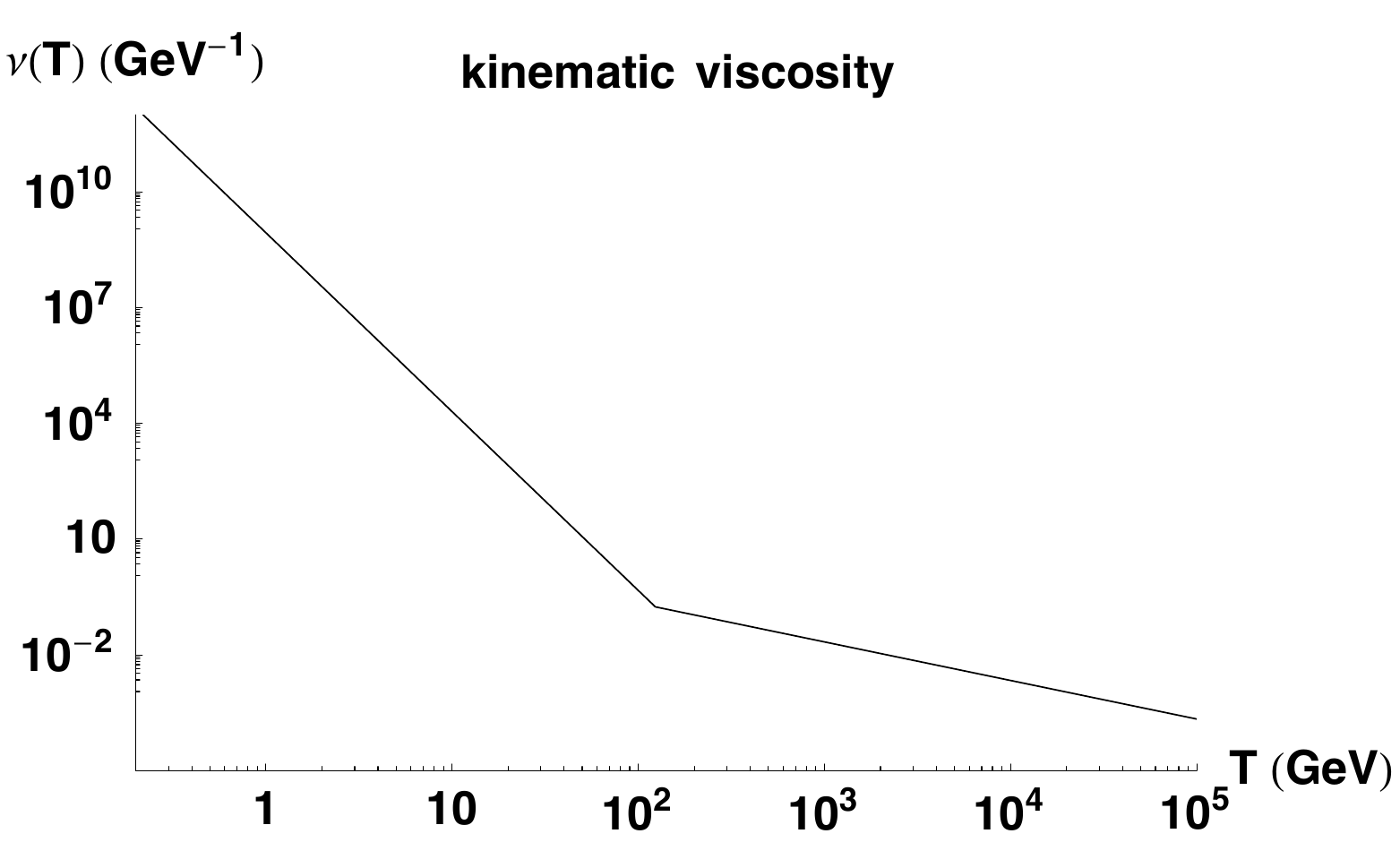}
\caption{\label{viscosity} \small Evolution with temperature of the kinematic viscosity $\nu(T)$.}
\end{center}
\end{figure}

After EW symmetry breaking, neutrino interactions are suppressed by a factor $(T/M_W)^4$. In this regime, neutrinos have the longest mean free path and dominate the viscosity. We use \cite{Heckler:1993nc}
\be
\ell_{\rm mfp}\approx (3 G_F^2 T^5)^{-1}\,,
\ee
leading to 
\be\label{e:nu100+}
\nu ( T\lesssim 100 \mbox{ GeV}) \approx 4.9 \times 10^8 \ \frac{\mbox{GeV}^4}{T^5}\,.
\ee
At temperatures smaller than $100$ MeV, after the QCD phase transition, the 
particle content changes and consequently  the neutrino mean free path 
increases to
\be
\ell_{\rm mfp}\approx \frac{10}{9} (G_F^2 T^5)^{-1}
\ee
leading to 
\be\label{e:nu100-}
\nu ( T\lesssim 100 \mbox{ MeV}) \approx 1.6 \times 10^9 \ \frac{\mbox{GeV}^4}{T^5}\,.
\ee
At temperatures above the EW phase transition, neutrino interactions are no longer suppressed. The shear viscosity is dominated by right handed lepton transport and given by \cite{Arnold:2000dr}
\be
\bar{\eta} \approx \left(\frac{5}{2}\right)^3\zeta(5)^2\left(\frac{12}{\pi}\right)^5\frac{3/2}{9\pi^2+224 (5+1/2)} \frac{T^3}{g'^4\log g'^{-1}}
\ee
where $g'$ is the hypercharge coupling. This leads to
\be
\nu ( T\gtrsim 100 \mbox{ GeV}) \approx  \frac{21.6}{T}\,.
\ee
The evolution of $\nu$ with temperature is plotted in Fig.\ref{viscosity}.
The neutrinos remain the relevant particles controlling the viscosity until 
they decouple at $T\sim 1.4$ MeV, after which photons take over. Even if there 
was some source of turbulence after the EW phase transition, turbulence 
is expected to terminate anyway around 1 MeV. Indeed,  $e^+e^-$ annihilation 
 reduces the plasma electron population and  
increases the photon diffusion length hence also the kinematic viscosity, 
leading to a decrease of the Reynolds number below one. 

Since we have only found non-relativistic derivations in the 
literature~\cite{LLk,MHDturb,BS}, let us estimate here in some detail the 
magnetic diffusivity and the magnetic Prandl number for relativistic electrons
in the cosmic plasma with temperatures $1$ MeV $< T< 100$ GeV,
\be\label{Ae:Pm}
{\rm P_m}(T) \equiv \frac{{\rm R_m}(L,T)}{\Re(L,T)} = \frac{\nu(T)}{\mu(T)} \, .
\ee
We want to determine ${\rm P_m}(T)$ when the electrons are relativistic and 
their dominant interactions are electromagnetic.

To determine the magnetic diffusivity $\mu(T)$ we derive an expression for the 
conductivity $\si(T)$.  The Lorentz force acting on an electron is
$$m_e\frac{du^\mu}{d\tau} = eF^{\mu\nu}u_\nu \,.$$ 
If we average this equation over a fluid element containing many electrons, the magnetic field
term is sub-dominant. Even though the electrons are highly relativistic, the average fluid velocity
is small. Furthermore $\gamma =1/\sqrt{1-v_e^2} \simeq T/m_e$ is nearly constant and we may
neglect the contribution $d\ga/d\tau$ from $du^i/d\tau = d(\ga v^i)/d\tau$ above. 
With $d\tau = \ga^{-1}dt = (m_e/T)dt $, this yields the following equation for
the mean velocity of the electron fluid: 
$$ \frac{d\bv}{dt} = \frac{e}{T}\,{\bf E} \, .$$
If we denote the collision time for the electrons by $t_c$, they can 
acquire velocities of the order $\bv \simeq \frac{e}{T}\,{\bf E}\,t_c$ 
between successive collisions. Hence the current is 
$${\bf J} \simeq en_e{\bf v} \simeq t_c\frac{e^2n_e}{T}{\bf E}\equiv 
\si {\bf E}$$ 
so that the conductivity  becomes
$$ \si =  t_c\frac{e^2n_e}{T} \,. $$
We now derive an estimate for $t_c$ from Coulomb interactions.
For a strong collision between the electron and another charged particle we
need an impact parameter $b$ such that $e^2/b > E_e \simeq T$. Hence the
cross section becomes $\si_t \sim \pi b^2 \simeq \pi e^4/ T^2$ (this simple 
argumentation neglects the Coulomb logarithms which enhance the cross section
by $\ln(1/\al_{\min})$ where $\al_{\min}$ is the minimal deflection 
angle~\cite{LLk}). With $v_e=1$ the time between collisions is therefore
$ t_c= 1/(\si_tn_e) \simeq T^2/(\pi  e^4n_e )$ and
\be\label{e:si}
\si \simeq \frac{T}{\pi e^2} \,. 
\ee
Note that this result is independent of the electron density. This is 
physically sensible as $n_e$ enhances the current on the one hand but it 
reduces in the same way the collision time.

With (\ref{e:si}) we obtain for the magnetic diffusivity
\be\label{e:mu}
\mu(T) \equiv \frac{1}{4\pi\si} \simeq \frac{e^2}{4T} \simeq 
\frac{10^{-3}}{T} \, .
 \ee
Inserting the kinematic viscosity from Eqs.~(\ref{e:nu100+}) or (\ref{e:nu100-}) 
we obtain for the Prandl number
\be\label{e:Papp}
{\rm P_m} =\frac{\nu}{\mu}\simeq 10^{12}\left(\frac{\rm GeV}{T}\right)^4 \, .
\ee
This number is  larger than $1$ for all temperatures $1$ MeV$<T<100$ GeV 
where the derivation applies. Hence currents and magnetic fields 
can develop and we are in the regime where MHD turbulence applies.

\section{The unequal time anisotropic stress power spectrum \label{app:turb}}

As mentioned in the main text, the unequal time anisotropic stress power 
spectrum $\Pi(k,t_1,t_2)$ is given by the convolution of the unequal time 
source power spectrum. The details of the derivation (for the formally 
identical case of a magnetic field) can be found in Ref.~\cite{GW1mag}; 
here we just give its main steps. One starts with Eq.~(\ref{Pifour}).
Wick's theorem gives
\bea
\lefteqn{\langle v^i({\bf k})v^{*j}({\bf q}) v^n({\bf s})v^{*m}({\bf p}
)\rangle   = ~~~~~~~~~~} \nonumber \\  &&
~~~~~~~~~~\langle v^i({\bf k})v^{*j}({\bf q})\rangle
\langle v^n({\bf s})v^{*m}({\bf p})\rangle + \nonumber \\ &&
~~~~~~~~~~ \langle v^i({\bf k})v^{n}({\bf s})
\rangle\langle v^{*j}({\bf q})v^{*m}({\bf p})\rangle + \nonumber \\ &&
~~~~~~~~~~ \langle v^i({\bf k})v^{*m}({\bf p})\rangle
\langle v^n({\bf s})v^{*j}({\bf q})\rangle ~. \label{corV2}
\eea
This has to be inserted in Eq.~(\ref{Pifour}) together with the definition of 
the velocity power spectrum (\ref{velunequaltime}). Applying the projection 
operator gives the angular dependence,
\begin{eqnarray}
& &{\mathcal P}^{abcd}(\hat\bk)[(\delta^{ac}-\hat{q}^{a}\hat{q}^c)
(\delta^{bd}-(\widehat{\bk-\bq})^{b}(\widehat{\bk-\bq})^d) \nonumber \\
& & ~~ \qquad +(\delta^{ad}-\hat{q}^{a}\hat{q}^d)
(\delta^{bc}-(\widehat{\bk-\bq})^{b}(\widehat{\bk-\bq})^c)]= \nonumber \\
& &1+(\hat{\bk}\!\cdot\!(\widehat{\bk\!-\!\bq}))^2+
(\hat{\bk}\!\cdot\!\hat{\bq})^2+
(\hat{\bk}\!\cdot\!\hat{\bq})^2
(\hat{\bk}\!\cdot\!(\widehat{\bk\!-\!\bq}))^2\,.
\label{alg}
\end{eqnarray}
Comparing then with Eq.~(\ref{anisostress}), one arrives at the result:
\be
\Pi_v(k,t_1,t_2)=\int d^3p P_v(p,t_1,t_2)P_v(|\bk-\bp|,t_1,t_2)
(1+\gamma^2)(1+\beta^2)\,,
\label{a:Pispectrumint}
\ee
with $\gamma=\hat{\bk}\cdot\hat{\bp}$, $\beta=\hat{\bk}\cdot\widehat{\bk-\bp}$.

\end{document}